%% file: paper1_21aug.tex
\documentclass[galley,usenatbib]{mn2e}
\usepackage{ textcomp }

\include{journal_abbv}
\usepackage{graphicx}	
\usepackage{amsmath}	
\usepackage{amssymb}	
\usepackage{multicol}        
\usepackage{bm}		
\usepackage{pdflscape}	
\usepackage{float}
\usepackage{subcaption}
\usepackage{longtable}
\usepackage{journal}
\hbadness=\maxdimen
\vbadness=\maxdimen

\title[LMC star clusters]
  {STAR CLUSTERS IN THE MAGELLANIC \\
CLOUDS-1: PARAMETERISATION AND CLASSIFICATION OF 1072 CLUSTERS IN THE LMC}
\author[P. K. Nayak et al.]
  {P. K.~Nayak,$^1$\thanks{E-mail: prasanta@iiap.res.in}
  A.~Subramaniam,$^1$ S.~Choudhury,$^{1,2}$ G.~Indu,$^1$
  Ram~Sagar$^1$ \\
  $^1$Indian Institute of Astrophysics, Bangalore 560034, India\\
  $^2$Indian Institute of Science, Bangalore 560012, India}

\def\LaTeX{L\kern-.36em\raise.3ex\hbox{a}\kern-.15em
    T\kern-.1667em\lower.7ex\hbox{E}\kern-.125emX}

\begin{document}

\label{firstpage}

\maketitle

\begin{abstract}
We have introduced a semi-automated quantitative method to estimate the age and reddening of 1072 star clusters in the Large Magellanic Cloud (LMC) using the Optical Gravitational Lensing Experiment (OGLE) III survey data. This study brings out 308 newly parameterised clusters. 
In a first of its kind, the LMC clusters are classified into groups based on richness/mass as very poor, poor, moderate and rich clusters, similar to the classification scheme of open clusters in the Galaxy. A major cluster formation episode is found to happen at 125$\pm$25 Myr in the inner LMC. The bar region of the LMC appears prominently in the age range 60 - 250 Myr and is found to have a relatively higher concentration of poor and moderate clusters. The eastern and the western ends of the bar are found to form clusters initially, which later propagates to the central part. We demonstrate that there is a significant difference in the distribution of clusters as a function of mass, using a movie based on the propagation (in space and time) of cluster formation in various groups. The importance of including 
the low mass clusters in the cluster formation history is demonstrated. The catalog with parameters, classification, and cleaned and isochrone fitted CMDs of 1072 clusters, which are available as online material, can be further used to understand the hierarchical formation of clusters in selected regions of the LMC.
\end{abstract}


\section{Introduction}

Stars in a cluster are generally assumed to be born from the same material, at about same age and also located at the same distance.  Since  their characteristics depend upon the environment of the host galaxy, study of star clusters helps us to understand star formation history (SFH), recent star formation and the formation and evolution of the host galaxy. 

The Large Magellanic Cloud (LMC), one of the neighbouring galaxies to the Milky way (MW) appears almost face on to us \citep{smitha2013} and situated at the distance of $\sim$ 50 kpc, hosts a large number of star clusters. Due the proximity of the LMC one can resolve the individual  members of the star clusters with ground based telescopes under good seeing conditions. 

The most recent and extensive catalogue of star clusters in the LMC is given by \citet{Bica2008} (hereafter B08). It  provides central coordinates, radii and position angles for $\sim$ 3000 star clusters. The parameters of some of these clusters have been estimated by a number of investigators. 
\citet*{glatt2010} (hereafter G10), using the Magellanic Cloud Photometric Survey (MCPS) \citep{zaritsky2002,zaritsky2004} data, presented ages and luminosities of 1194 star clusters and associations in the LMC. 
\citet{pu2000} (hereafter PU00) estimated ages and reddening of 600 star clusters in the central LMC, using the Optical Gravitational Lensing Experiment II (OGLE II) data \citep*{udalski1997_ogleII}. Using Washington photometric data, parameters of 277 LMC star clusters were estimated by \citet{palma2016_catalog}, \citet{chou2015}, \citet{Piatti2002_catalog,piatti2003a,piatti2003b,piatti2009b_catalog}, \citet{Piatti2011_catalog,piatti2012_catalog,piatti2014_catalog}. Recently \citet{piatti2014_vmcXII,piatti2015_vmcXVI} estimated parameters of 378 star clusters using near-IR data from the VMC survey.
Most of the well studied clusters in the LMC are rich star clusters, which stand out from the field region due their high stellar density. A large fraction of star clusters are either not well-studied or are unstudied due to the poor nature of the cluster \citep{chou2015}. A more precise determination of cluster parameters of larger sample is required to understand the star cluster formation history of the LMC.

The LMC has an off-centred bar that appears as an over density in young and
old stellar populations \citep{cioni2000,niko2004,smitha2013,vankal2014,zhao2000_lmcbar,vander} 
as well as in the numerical models of the bar
\citep{bekki2009,besla2012}. A comprehensive sample of star clusters with parameters will be helpful to identify the episodes of star formation in the bar. In the LMC, G10 found the cluster formation to peak at 125 and 800 Myr, whereas PU00 found the cluster formation to peak at 7, 125 and 800 Myr. 
So the star clusters could also trace any propagating star formation in the bar region during the burst of star formation in the 100-125 Myr age range.  Recently,  \citet{jacy2016} analysed OGLE IV Cepheids and found that the central part of the bar has the youngest cepheids with relatively older cepheids in the eastern and
western parts.  \citet{piatti2015_vmcXVI} studied the VMC tiles in the central LMC and found that the outer most regions of the bar
experienced star formation earlier and the central regions have younger clusters. The authors also found that the 30 Dor region is dominated by recent cluster formation.

The LMC is well known for the presence of rich star clusters, which are a few 100 Myr old. Recent studies have shown that the LMC also
hosts a good number of poor star clusters, which are similar to the open star clusters of our Galaxy \citep{piatti2012_catalog,chou2015}. 
\citet{baum2013} showed that the mass of the star clusters in the LMC cover a large range. They studied the mass function of clusters massive than 5000 M$_\odot$, though the cluster mass range extended down to a few 100 M$_\odot$ in the low mass limit. The cluster mass function of LMC clusters were studied by \citet{grijs2013}, where they do not find any evidence of early disruption of young clusters more massive than 10$^3$ M$_\odot$. Thus, the star clusters in the LMC have a wide range in mass and it is essential to classify star clusters into groups corresponding to their mass.
\citet{kontizas} had classified clusters in the LMC as Compact (C) or Loose (L) based on their appearance in the photographic plates, which is not related to any physical parameters of the clusters. \citet{searle} had classified 61 star clusters in the Magellanic Clouds (MCs) as type I to VII based on their integrated spectra using four colour photometry of integrated light. Even though there are many studies to understand the cluster mass function of star clusters in the LMC, there has been no attempt to classify each star cluster and group them according to their mass/strength, so far. On the other hand, the classification of Open Clusters (OCs) in the Galaxy dates back to 1930, \citep{trump30}. 
Trumpler$'$s classification made use of the full information obtained from a single photograph of a cluster: degree of central concentration of stars, the range in luminosity of the members, the number of stars contained in the cluster and necessary conspicuous properties \citep{ruprecht66}.  
 This system known as Trumpler$'$s system of classification is used to homogeneously classify most of the open clusters.

The parameters of a star cluster are estimated from its colour-magnitude diagram (CMD). 
Generally, a visual fitting of the isochrone is performed to the cluster CMD  
to estimate the reddening and age of the cluster.  
The visual fitting of the cluster CMD can be performed effectively for a sample of a small number of clusters. When the cluster sample exceeds a few hundred, this method is not very efficient. Also, to estimate the parameters in a self-consistent way, it is ideal to have a quantitative method in order to eliminate systematic errors.

The aims of this study are multifold, (1) to increase the sample of well studied clusters and to a consistently estimate the parameters like age and reddening of already identified star clusters of the LMC utilising available photometric survey data, (2) to develop a quantitative method to estimate cluster parameters and (3) to classify clusters based on their richness/mass.

The remaining part of the paper is arranged as follows: Section 2 deals with Data, followed by analysis in section 3. The cluster classification scheme is presented in section 4 and the error estimation is presented in section 5. The results and discussion are presented in section 6, followed by the summary in section 7.

\section{Data}

In this study, we have identified star clusters of the B08 catalog in the Optical Gravitational Lensing Experiment III (OGLE III \citep{udalski2008_ogleIII}) observed region and characterised them using the V,I photometric data. 
The OGLE III observations were carried out at Las Campanas Observatory, operated by the Carnegie Institution of Washington, with 1.3-m Warsaw telescope equipped with the second generation mosaic camera consisting of eight SITe 2048 $\times$ 4096 CCD detectors with pixel size of 15 $\mu$m, which corresponds to 0.26 \arcsec/pixel scale. The OGLE III survey observed an area of about 40 square degrees around the LMC centre, and presented the V and I magnitudes of about 35 million stars in the LMC fields \citep{udalski2008_ogleIII}. For our study, we have considered stars with photometric errors in V and I bands $\le$ 0.15 mag. We note that this limiting error is in general found for stars near the fainter end of the MS, whereas the stars near the turn-off, which are used to estimate the cluster parameters have relatively less photometric error depending on the turn-off magnitude. For the stars brighter than 19 magnitude, the photometric error is found to be $\le$ 0.05 mag.

The OGLE III data cover the central region of the LMC (including the bar region) but does not cover the northern star forming regions. This data is therefore ideal to study the cluster formation in the central LMC, particularly in the bar. This study is complementary to that of G10, where most of the clusters studies are located outside the bar region. PU00 used the OGLE II survey data and studied clusters located in the bar region. As the resolution of OGLE III data is better than the OGLE II data, this study will be able to improve the estimates of PU00, particularly in the bar region which has maximum crowding. The OGLE III data is useful to characterise young star clusters, with ages $<$ 1 Gyr. This data will be useful to detect the bursts of cluster formation, which are linked to the interaction between the LMC and the SMC.

\section{Analysis}
\subsection{Cluster sample}

As mentioned earlier, we have started from the catalog by B08 and identified 1765 star clusters located well within the OGLE III observed field. 
 We have used the major and minor axes of each cluster listed in the catalog by B08 to calculate their radius, which is defined as \textonequarter(major + minor axis).
The  estimated cluster radii are found to range from 0.20\arcmin to 1.75\arcmin on the sky, with physical sizes corresponding to $\sim$2.9 to 25.4 pc respectively. We extracted data of the cluster regions (stars within the cluster radius) from the OGLE III field along with a few arcmin field region around them, depending upon their radii.  The estimated number of stars within the cluster radius is denoted as ($n_c$).

A cluster region is observed as a density enhanced region with respect to the surrounding field, and consists of cluster members as well as field stars. The LMC is known to host very rich as well as poor clusters, located in a range of environments of stellar density.
 The fundamental feature of a cluster which is used to estimate the reddening, age and distance is the main-sequence (MS) and the location of turn-off in the CMD. The field star removal is necessary to define the cluster sequence and this will depend on the field star density and its variation in the vicinity of the cluster. Therefore, we restricted our analysis to
star clusters, where one can reliably remove field stars and identify the desired cluster features. We describe the adopted method below. As a first step, we identified clusters which are located in regions with large variation
in the field star density. In order to measure the variation in field star distribution surrounding the cluster field, we chose four annular field regions (each of equal area as the cluster region) around each cluster. The inner radii of the four annular field regions are chosen as 0.5$\arcmin$, 1.0$\arcmin$, 1.5$\arcmin$ and 2.0$\arcmin$ larger than the cluster radius. Number of stars in each field region is counted.
The number of field stars contaminating the corresponding cluster is estimated as the average number of stars of four field regions ($n_f$).  Standard deviation ($\sigma_f$) about the average indicates the variation in the stellar field counts. The number of cluster stars ($n_m$) or the strength of the cluster is then defined as $n_m = n_c - n_f$. 

\begin{figure}
 \includegraphics[width=\columnwidth]{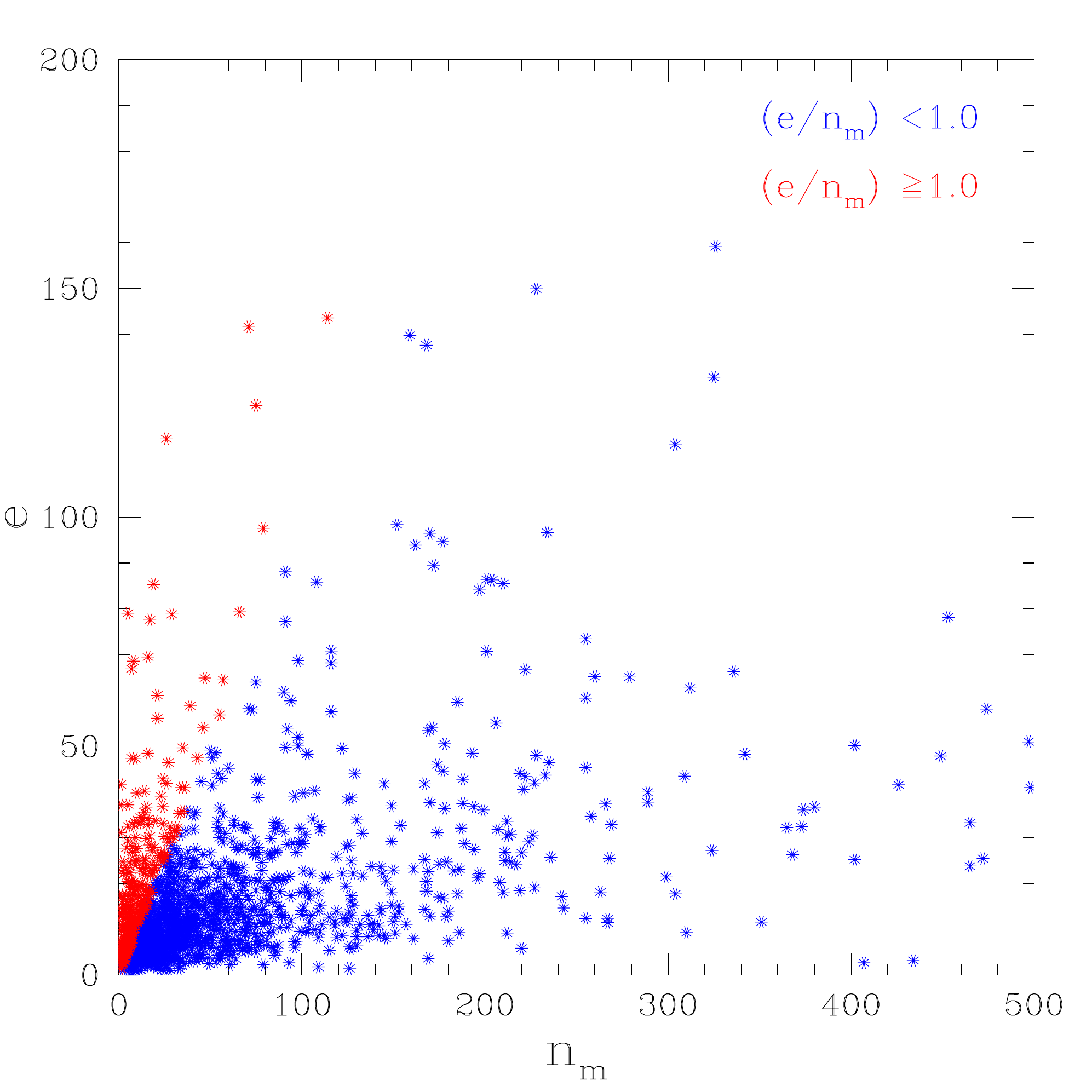}
 \caption{The distribution of error ($e$) with respect to cluster member strength ($n_m$). The clusters with $e/n_m>1$ are denoted as red points, whereas, the blue points denote the clusters with $e/n_m\le1$.}
 \label{error}
\end{figure}

We have then separated star clusters on the basis of variation in the field star distribution. We have excluded clusters which are
embedded in fields suffering from large dispersion in the field star count with respect to the average, around the cluster. We have also excluded clusters, where the value of dispersion in the field star count is equal to or greater than the number of stars in the cluster. 
In order to achieve this, we have used the following criteria : \\
	(i) We estimated the fractional standard deviation as $\sigma_f/n_f$, to quantify the variation in field star count. We excluded clusters where variation in the field stars count is greater than or equal to 50$\%$ of the average count, i.e. $\sigma_f/n_f$ $\ge$ 0.5. With this criteria, we excluded 48 clusters.\\
	(ii) The variation in the field star counts will propagate as an error when we estimate the strength of the cluster. The error associated with the estimation of $n_m$ can be defined as :\\
\begin{equation}
		e= |(n_c - (n_f + \sigma_f)) - (n_c - (n_f - \sigma_f))|,
\label{error_relation}
\end{equation}
which is basically the difference between the maximum and minimum value of $n_m$, for $\sigma_f$ deviation in the field star distribution. For crowded field regions, there is a possibility that $\sigma$$_f$ is high, so will be the value of $e$. In order
to remove clusters, where the error itself is greater than the number of stars in the cluster, 
we calculated the fractional error as $e/n_m$ and excluded cases with $e/n_m$ $\ge$ 1. Based on this criteria, we excluded 310 clusters from our sample. Figure \ref{error} shows a plot between $e$ and $n_m$. The cases with $e/n_m$ $\ge$ 1 are shown in red, whereas the cases with $e/n_m$ $<$1 are shown as blue points. It is seen that clusters which are relatively poor, with $n_m$ $\la$ 30, have $e/n_m$ $\ge$ 1. 

The number of clusters remaining in the sample, after implementing the above two cut-off criteria is 1407. Out of these clusters, 46 are relatively rich clusters ($n_m$ $>$ 400) and we have excluded them from our analysis as they are already well studied using better data. Thus we proceeded with the sample of 1361 clusters for further analysis.

\subsection{Removal of field star contamination from cluster CMDs}
 
 We constructed (V, V $-$ I) CMDs for cluster and field regions to compare and remove field star contamination from the cluster region, using a statistical process. The annular field chosen to carry out this process is the one with inner radius 0.5$\arcmin$ larger than cluster radius.
 The field stars within the cluster region are then removed by considering each star in the field CMD and finding its nearest counterpart in the cluster CMD. We
 considered a grid of [magnitude, color] bins with different sizes, starting with [$\Delta$V , $\Delta$(V $-$ I)] = [0.02, 0.01] up to a maximum of [0.5, 0.25], where the units are in magnitude. This procedure is repeated for all the clusters. The cleaned cluster CMDs show prominent cluster features with minimum unavoidable field star contamination. The similar technique has been used earlier by \citet{chou2015}.

\subsection{Semi-automated quantitative method}
		Now we have the CMDs of 1361 star clusters with field stars removed and ready for the estimation of cluster parameters. The age and reddening of star clusters are generally estimated by the visual fitting of isochrones to the main sequence turn-off (MSTO) of cluster CMDs. 
However, visual fitting of isochrones to 1361 clusters is not only a laborious task, but also produces inconsistently estimated parameters across the sample. For the first time, we have developed a semi-automated quantitative method to estimate cluster parameters, so that we can not only estimate the parameters accurately and consistently, but also quantify errors. We adopted a two step process for the estimation of cluster parameters. In the first step, we developed a quantitative automated method to estimate the age and reddening (as describe below). This method is applied to all clusters to estimate the reddening and age. In the second step, we plotted the isochrones on each of the 1361 cluster CMDs, for the estimated age and corrected for the estimated reddening. All the CMDs were visually checked and corrected for any error in the fit. The second step is used to check and correct for improper estimation of parameters by the automated method. Thus, the method we developed in this study could be termed as a semi-automated quantitative method. We describe the development of the quantitative automated part below.

The processes involved in this method are : \\
(a) Identifying the MS in the cleaned cluster CMD and constructing the MS luminosity function (MSLF).\\
(b) Identifying the MS Turn-off (MSTO) and estimating the corresponding apparent magnitude and colour.\\
(c) Estimating the reddening from the (V$-$I) colour of the MSTO.\\
(d) Estimating the absolute magnitude of the MSTO after correcting for reddening and distance.\\
(e) Estimating the age using age-magnitude relation derived using \citet{marigo2008} (hereafter M08) isochrones.\\

The above steps are described in detail below.

		(a) We consider stars brighter than 21 mag in V and bluer than 0.5 mag in (V $-$ I) colour as the Main sequence (MS) stars. To construct the MSLF, the magnitude axis is binned with a bin size of 0.2 mag. The brightest bin with a minimum number of stars ($\eta$) is identified as the bin corresponding to the MSTO. The mean V magnitude of the brightest bin is considered as turn-off V (V$_{TO}$) magnitude. The MSTO bin (which is likely to be the brightest bin of the MSLF) needs to be identified from the MSLF using statistically significant value of $\eta$ so that 
it excludes blue supergiants. The value of $\eta$ in a cluster will depend mainly on richness as well as age of the cluster.
Two clusters which are similar in richness, but with different age will have different MSTO bin with different $\eta$. The MSTO bin will be less populated for a younger cluster than the older one with similar richness. Two clusters with same age but different number of cluster stars will also have different values of $\eta$ for their MSTO bin. A very similar idea was used by \citet{indu2011} to identify the MSTO of field regions in the Magellanic Clouds. As mentioned by these authors, the identified bin and the number of stars in the bin are dependent on the richness/area of the field. 
Here, the known parameter is the richness (total number of cluster stars) of the cluster, as we are yet to estimate their age. Therefore, we have grouped the clusters according to their strength and fixed $\eta$ value for each group. We have discussed the groups and the selection of $\eta$ for each group in section 3.4. 

		(b) The mean V magnitude of the MSTO bin is considered as turn-off V magnitude of cluster (V$_{TO}$). Once we have identified the V$_{TO}$ the next task is to estimate the colour of the MSTO. The colour of the MSTO can be identified as the peak in colour distribution near the MSTO.
To estimate the peak colour of the MSTO,  a strip parallel to colour axis with a width of 0.6 mag about V$_{TO}$ is considered (V$_T$$_O$ + 0.4 mag to V$_T$$_O$ - 0.2 mag). This is to ensure that we have a statistically significant number of stars near the MSTO. For the clusters with $n_m\le$ 100, a width of 0.8 mag is considered (given by V$_T$$_O$ + 0.6 mag to V$_T$$_O$ $-$ 0.2 mag). The choice for width of the strip does not affect the position of the peak colour, as the isochrones for younger ages are almost vertical to the colour axis near the MSTO. This strip is binned in colour with a bin size of 0.1 mag to estimate the distribution of stars along the colour axis. The distribution is found to have a unique peak (in most of the cases) with asymmetric wings. The mean colour of the bin corresponding to this unique peak  
is chosen as the apparent color, (V $-$ I)$_{app}$, of the MSTO.

		(c) The reddening of the cluster is defined as the difference between the apparent and absolute colour of the MSTO. 
To begin with, we have adopted A$_V$ = 0.55 \citep{zaritsky2004} and distance modulus (DM) = 18.50 $\pm$ 0.10 \citep{saha2010} for the LMC.
If M$_V$ is the absolute magnitude of the MSTO, then assuming a distance modulus and an average value of extinction (A$_V$) for the cluster, the apparent magnitude (V$_{TO}$) is related to M$_V$ as:
\begin{equation}
		M_V = V_{TO} - DM - A_V,
\label{mv_vto_relation}
\end{equation}

The estimated value of M$_V$ is cross-matched with the absolute V magnitude of MSTOs from the isochrones table of M08, with Z = 0.008 \citep{piagei2013} for the LMC. The (V $-$ I) colour corresponding to the closest match of absolute MSTO V magnitude gives the absolute colour for the MSTO, (V $-$ I)$_0$. The reddening (E(V $-$ I)) for the the cluster is then given as:
\begin{equation}
		E(V - I) = (V - I)_{app} - (V - I)_0. 
\label{evi_relation}
\end{equation}
The extinction for the cluster region is estimated as, A$_V$ = 2.48$\times$E(V $-$ I) \citep{niko2004}.
 The extinction corrected value of M$_V$ of the MSTO is then calculated again by using this value of A$_V$ in
Equation \ref{mv_vto_relation}. This process was iterated a couple of times and it was found that the values estimated do not change with respect
to the values obtained in the first iteration.
The method used here is similar to that adopted by \citet{indu2011}, for estimating the reddening
of field regions.

\begin{figure}
\includegraphics[width=\columnwidth]{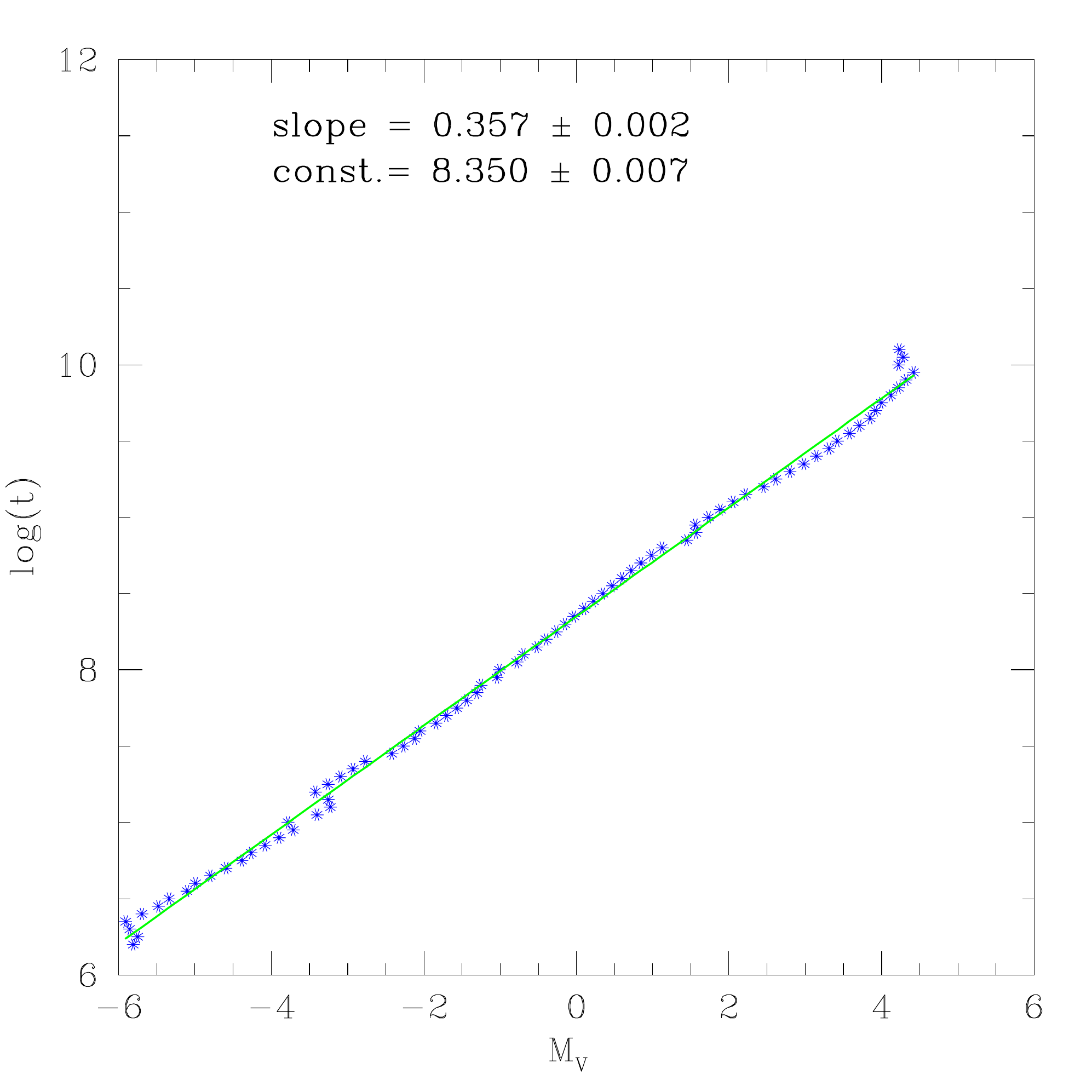}
\caption{The relation between the absolute turn-off V magnitude (M$_V$) and age, within the range log(t) = 6.2 to 10.2 for M08 isochrones. A straight line (green) fitted through the points is also shown.}
\label{log(age)-Mv}
\end{figure}

		(d) Figure \ref{log(age)-Mv} shows a plot between the absolute magnitude M$_V$ of the MSTO and their corresponding ages (log(t)) for M08 isochrones. The relation is found to be linear, and is given as:
\begin{equation}
		log(t) = 0.357(\pm0.002)M_V + 8.350(\pm0.006). 
\label{logt_mv_relation}
\end{equation}
The extinction corrected M$_V$ derived in step (c), is used in the above relation to estimate the ages of the clusters.

Thus, we have described a quantitative method to estimate the reddening and age of star clusters. This method primarily depends on the unique identification of the MSTO. As mentioned earlier, it is a function of age and richness of the cluster.

\subsection{Identification of MSTO and estimation of cluster parameters}
		In this section, we describe the method to identify the MSTO in the MSLF. The brightest bin with a minimum number of stars ($\eta$) is identified as the bin corresponding to the MSTO.  The bin corresponding to the MSTO will depend upon the selection of the minimum number criteria ($\eta$) for the brightest magnitude bin. As mentioned earlier, the value of $\eta$ will depend on the richness of the cluster. In subsection 3.1, we estimated the number of member stars in the cluster $n_m$, after subtracting the average field population. We have grouped clusters based on $n_m$, which is a measure of the richness of clusters and are shown in Table 1. We have performed this classification based on the number of cluster member stars which are brighter than V=21 mag, though it is directly linked to the mass of the clusters. 

\begin{figure}
\includegraphics[width=\columnwidth]{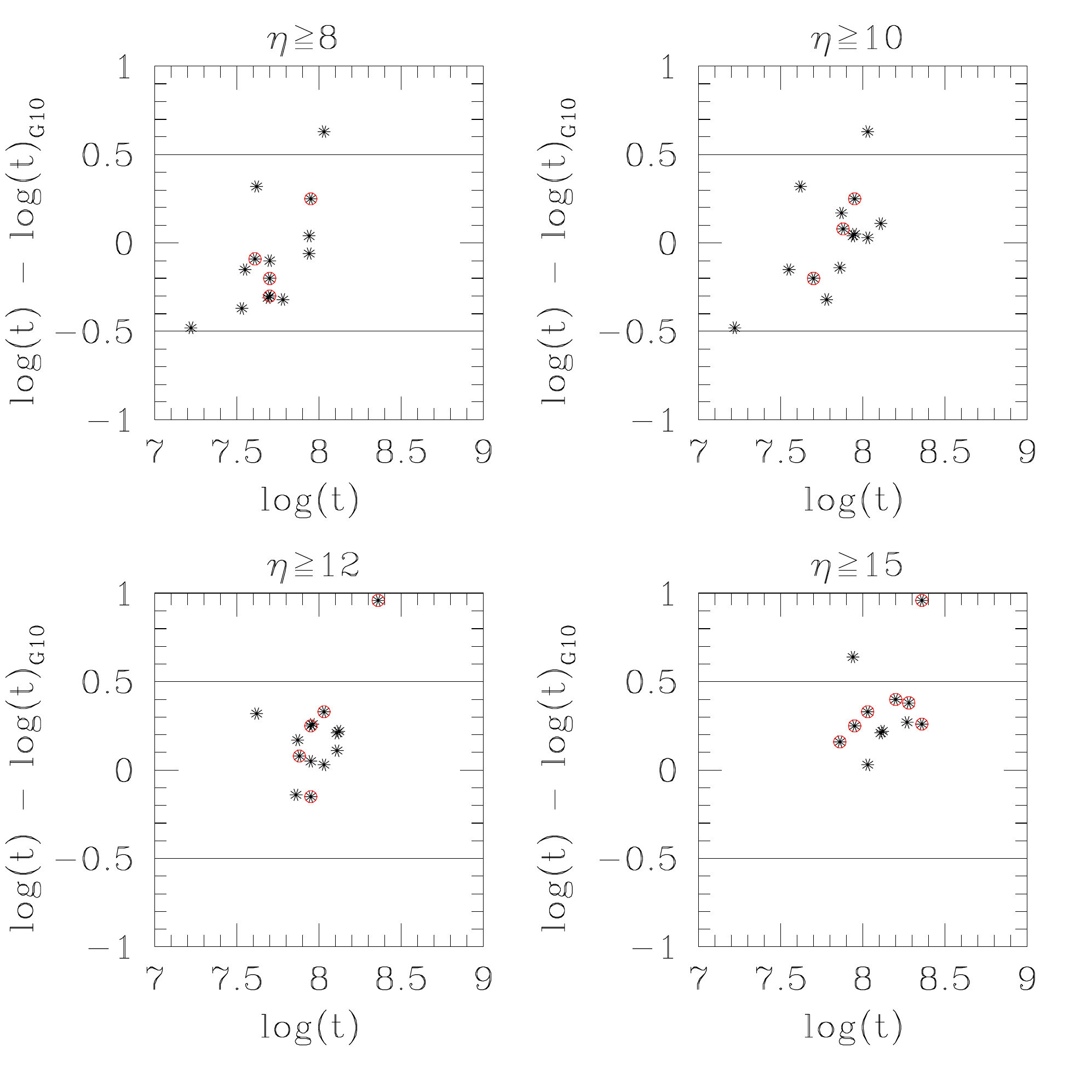}
\caption{Estimated age distribution of clusters for 200$<n_m\le300$ with respect to G10's estimation of age for different values of $\eta$. The clusters with $\delta$E(V $-$ I) $>$ 0.1 shown in Figure \ref{evidiff_le300} are marked as a red circle for corresponding values of $\eta$.}
\label{agediff_le300}
\end{figure}

\begin{figure}
\includegraphics[width=\columnwidth]{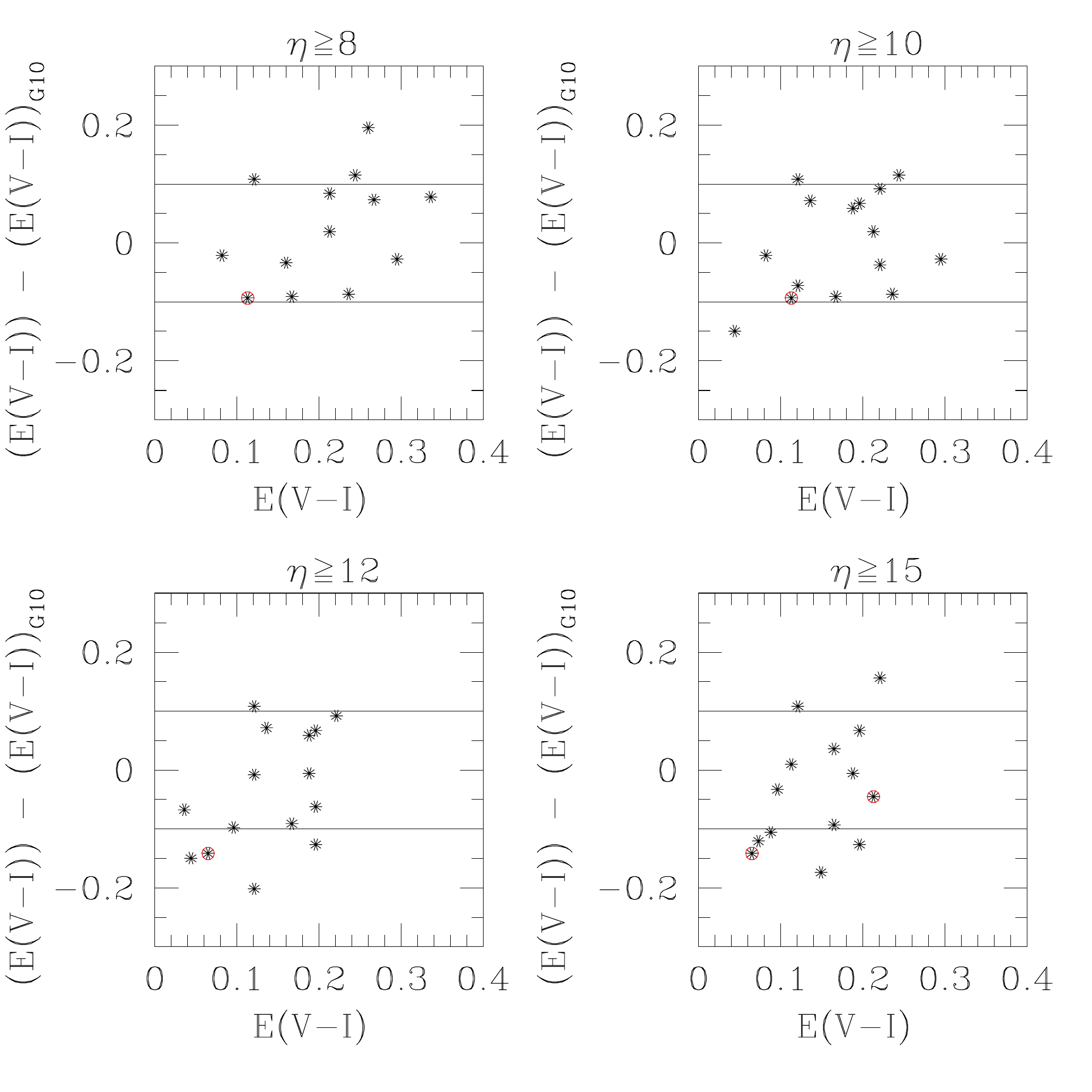}
\caption{Estimated reddening distribution of clusters for 200$<n_m\le300$ with respect to G10's estimation of ages for different value of $\eta$. The clusters with $\delta$log(age) $>$ 0.5 shown in Figure \ref{agediff_le300} are marked as a red circle for corresponding values of $\eta$.}
\label{evidiff_le300}
\end{figure}

The aim is to fix the value of $\eta$ for each group of clusters, classified according to their richness/mass so that a realistic MSTO and corresponding age of the cluster can be identified. If we fix the value low, then we would identify a brighter MSTO and estimate a younger age. And if we fix a higher value, we would then 
identify a fainter MSTO and estimate an older age for the same cluster. Thus, it is
important to fix an appropriate value for this parameter for each group. 
In order to calibrate the value of $\eta$, we identified clusters whose parameters (age and reddening) are already determined by G10, in each group. As we already have the field star corrected MSLF for these clusters, we estimated their age and
reddening using our method, for a range of $\eta$ values. The estimated ages and reddening (E(V $-$ I)) were compared with those estimated by G10 for the considered $\eta$ values. The maximum error in the age
estimation of G10 is 0.5 (in the log scale) and we choose this as the limit for comparing the age differences between the two estimates,
($\delta$log(t)). The maximum error in the reddening ($\delta$E(V $-$ I)) is 0.1 mag, corresponding to the error related to estimating colour peak in the MSTO magnitude bin.
The value of $\eta$, for each group of clusters is chosen such that there is least scatter in $\delta$log(t) and $\delta$E(V $-$ I) and the deviation is centered around the zero value. The values of $\eta$ determined for each group of clusters are listed in Table \ref{number_criteria}. 
In the table, we have also listed the number of clusters in each group whose ages and reddening have been compared with G10 (N$_{G10}$). 
In order to demonstrate the process, we discuss the analysis performed for one group of clusters (Group IV) with 200$<n_m\le$300. Figure \ref{agediff_le300} and Figure \ref{evidiff_le300} show plots of $\delta$log(t) and $\delta$E(V $-$ I) against age and reddening respectively, estimated by the above mentioned method. Four values of $\eta$ were used to estimate age and reddening. An inspection of  Figure \ref{agediff_le300} shows that the age estimation is better for $\eta\ge$10, as the deviation of $\delta$log(t) is almost symmetric about zero and has a spread over large age range with minimum error. For $\eta\ge$12, we observe clumping of data points, and for rest of the two criteria we estimated older age than G10 with relatively large error. 
In the Figure \ref{agediff_le300} the points with red circle indicates that the clusters have $\delta$E(V $-$ I) $>$ 0.1 for the corresponding $\eta$ value. In the Figure \ref{evidiff_le300} the clusters with $\delta$log(t) $>$ 0.5 for the corresponding $\eta$ value are marked as red circle. The comparison plots with $\eta$ value 10 show that the error in reddening value is not effecting the age estimation but for other $\eta$ values, it is found to affect the age estimation.

\begin{table}
\footnotesize
\centering
\caption{Grouping and classification of clusters based on their richness ($n_m$)/mass range (M$_c$) :}
\label{number_criteria}
\resizebox{84mm}{!}{
\begin{tabular}{ccccccc}
\hline
Group No. & Range of $n_m$  & $\eta$ & N$_{total}$ & N$_{G10}$ & Mass range (M$_\odot$) & Classification \\ \hline
 I        & 6$<n_m\le$30    & 2      & 438         & 149       & $<$ 800                & very poor  \\  \hline       
 II       & 30$<n_m\le$100  & 3      & 460         & 181       & 800 - 1700             & poor \\ \hline            
 III      & 100$<n_m\le$200 & 5      & 122         & 47        & 1700 - 3500            & moderate\\ \hline          
 IV       & 200$<n_m\le$300 & 10     & 43          & 15        & 3500 - 5000            & moderate\\ \hline
 V        & 300$<n_m\le$400 & 14     & 9           & 4         & $>$ 5000               & rich \\ \hline
\end{tabular}
}
\end{table}

We should point out that, initially we created 6 groups based on the value $n_m$ and estimated $\eta$ values for all the groups. It was found that the $\eta$ value was same for two groups, which we merged together to create the above five groups. These groups have
different values of $\eta$. The comparison of estimated ages with those estimated by G10 reveals that the dependency of age on the value of $\eta$ is not very large. This may be due to the fact that the age range considered in this study is not very large and most of the clusters are younger than 300 Myr.

\begin{figure}
\includegraphics[width=\columnwidth]{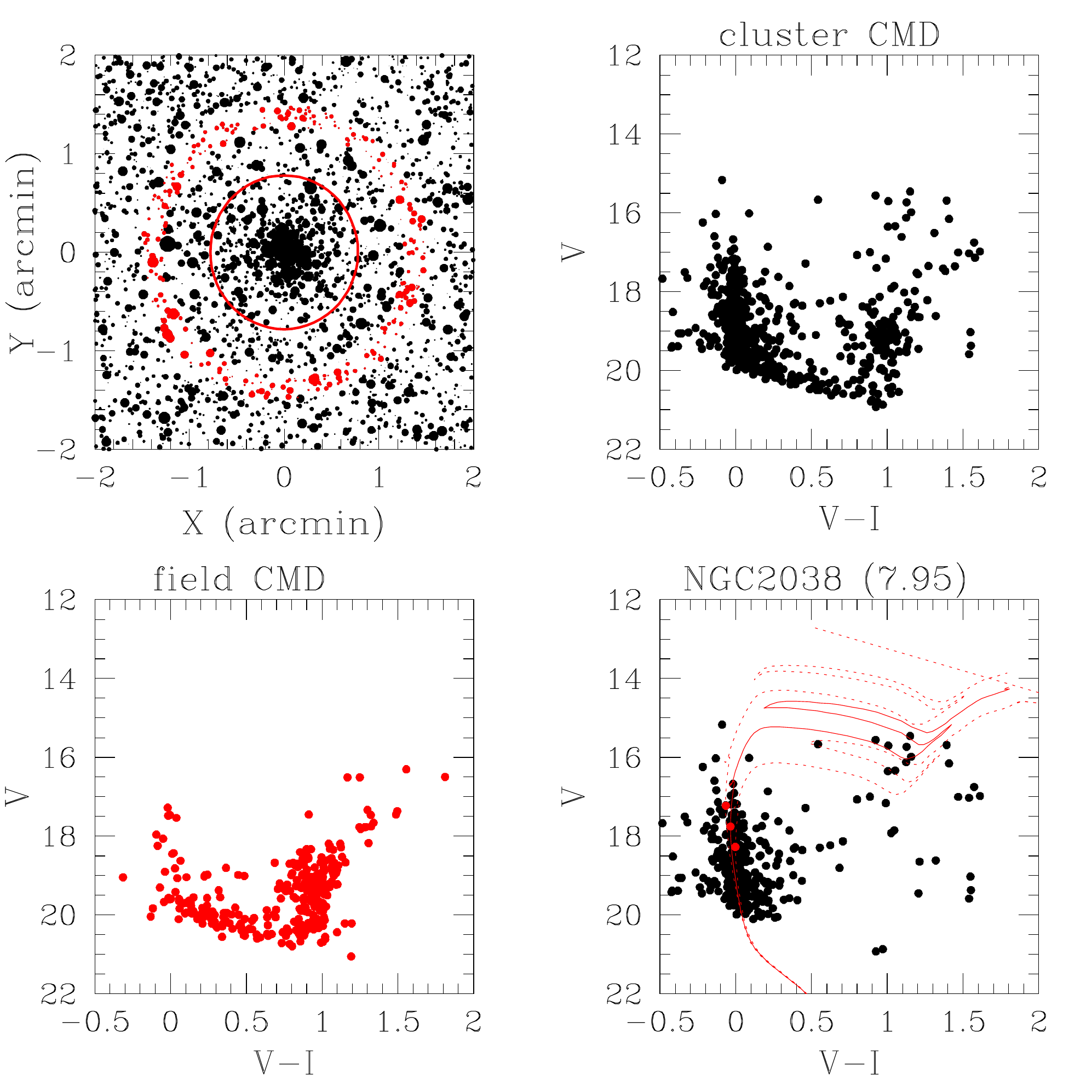}
\caption{Top left panel shows an example of a spatial plot of a cluster (NGC2038) and its surrounding field region, where the point sizes are proportional to brightness of stars in V magnitude. Cluster region is shown within the red circle whereas considered field stars are denoted as red points. The CMDs of cluster region (top right), annular field region (bottom left) and field star decontaminated cluster region (bottom right) are also shown. In the bottom right panel, isochrone of estimated age (red solid line) with an error of $\pm$0.2 (red dotted line) are over plotted on the cleaned CMD, where red dots are turn-off points of corresponding isochrones.}
\label{ngc2038_2730}
\end{figure}

We performed the above process for all the groups mentioned in the Table 1 and estimated the appropriate value of $\eta$ for each group. Now, we are ready to apply our quantitative method to all the clusters in our study. 
Since the limiting V magnitude of the OGLE III survey is $\sim$21 mag, the age and reddening of the clusters having V$_{TO}$ $\ge$ 19 mag could not be determined reliably using our method. There are 386 such clusters, which we have considered later.  We applied our method to 975 remaining clusters and estimated their age and reddening. A sample cluster is shown in Figure \ref{ngc2038_2730}, where the spatial location of the cluster along with the field region, CMD of all stars within the cluster radius, CMD of the field, and the CMD of the decontaminated cluster over plotted with the isochrone for the estimated age and corrected using the estimated reddening and DM are shown in four panels.

 Once we estimated the values
of age and reddening for each of the 975 clusters using the automated quantitative method, we proceed to the second step of our two step process. The second step is the manual check performed visually and is aimed at correcting the improper estimations of the automated method. The M08 isochrone for the estimated age (along with isochrones corresponding to age, log(t) $\pm$0.2) were overplotted on the cleaned cluster CMD after accounting for the reddening value.
All the plots are visually inspected for any mismatch between the plotted isochrone with respect to the MS and MSTO in the cluster CMD. After this visual inspection we found that, age and reddening of 752 (77.13$\%$) clusters are estimated reliably by the automated method, as the superimposed isochrones for the estimated age and reddening fit well to the cluster sequence. Whereas, a small correction in the reddening and age were found to be needed for 43 (4.41$\%$) and 67 (6.87$\%$) clusters respectively, to fit the isochrones in the cleaned CMD. Thus, we find that, the automated quantitative method is able to produce reliable fits for the majority of clusters in the sample. Including this small manual correction in the reddening and age, we are able to estimate cluster parameters of 862 clusters, which is 88.4\% of the sample studied here. It is worth mentioning that we changed the fitted values only if the required shift is more than the error in the estimation. That is, if there is a shift in reddening, which is less than 0.1 mag, we have not corrected it. And the same applies to the age estimation as well. We also tried to understand the reason for incorrect estimation of the parameters. We found that clusters with a flat profile for colour distribution, instead of a peak, had incorrect reddening estimation.  Similarly, clusters with gaps in the MS or clumpy MS, resulted in erroneous estimation of the MSTO. These are the cases which needed manual intervention to estimate reddening and age. 

For the remaining 113 out of 975 clusters, visual inspection revealed that the cleaned cluster CMDs have either large spread in the MS along colour axis or clumpy distribution along the MS. In some cases, there are big gaps in the MS along the magnitude axis. These features can be genuine or due to differential reddening, located in crowded fields or variation in the field star counts. 
The above problems may be due to the choice of a particular field region and hence we created three cleaned CMDs using three different field regions and plotted all the cleaned CMDs together. Stars which are not removed from the cluster CMD for at least two decontamination processes are considered as genuine cluster candidates. By this method, we were able to identify 22 clusters with prominent cluster features and corrected their parameters manually to get best fit of isochrones.

We examined the CMDs of the 386 clusters with V$_{TO}$ $>$ 19 mag and found that there are a few clusters with prominent clusters features or prominent features that can be extracted by using decontaminated CMDs using multiple field regions.
 We applied similar automated process to estimate the age and reddening of these clusters. We over 
plotted isochrones for estimated ages on cleaned CMDs of these clusters and manually corrected for ages and reddening wherever needed.
With this analysis we added 188 more clusters to the above mentioned 884 clusters with precisely estimated parameter. Thus we were able to determine ages and reddening for 1072 clusters, from the sample of 1361 clusters. The rest of the 289 clusters need more attention to estimate the cluster parameters. We plan to study these cluster in a separate future study.

\section{Classification of clusters and their mass range}
In the previous section we grouped star clusters based on the number of stars present in the cluster. We created groups as I, II, III, IV $\&$ V, which are arranged in the increasing strength of richness. As mentioned earlier, the grouping
is based on the number of stars present in the bin corresponding to the MSTO, which in turn is a function of the total number
of stars. Also, the number of stars in each cluster is directly related to the mass of the cluster. Basically, the fundamental parameter
which separates our sample into various groups is the mass of the cluster. 
Therefore, it will be interesting to estimate the mass range for the above mentioned groups. It is to be noted that the aim here is to give a physical sense to the above classification based primarily on the richness of the clusters.

As most of the clusters studied here are younger than 300 Myr, we can estimate the range of mass (M$_c$) occupied by these groups using the the $\eta$ parameters and assuming a typical age of 100 Myr. 
 We constructed synthetic CMD for the clusters using M08 isochrones, for a mass range of 0.1-15.0M$_\odot$. We assumed Salpeter$'$s mass function and also incorporated observational errors. In order to suppress statistical fluctuation due to the low value of $\eta$, we simulated cluster CMD for a large value of the total cluster number ($\sim$ 10$^6$). The synthesised LF is then scaled for the value of $\eta$ (for the MSTO bin) and the total mass of the cluster is estimated using the scaling factor. We have allowed for statistical error, while matching the value of $\eta$. Thus, we estimated the masses for the groups 
and classify the above groups as follows:\\ 
(1) Group I: This group has very less number of stars and has mass M$_c$ $<$ 800 M$_\odot$. Clusters in this group are classified as {\it very poor} clusters. This group has 438 clusters. \\
(2) Group II: This group has relatively less number of stars and has mass in the range M$_c$ $\sim$ 800 - 1700 M$_\odot$. Clusters in this group are classified as {\it poor} clusters. This group has 460 clusters. \\
(3) Groups III $\&$ IV: This group has relatively large number of stars and has the mass range M$_c$ $\sim$ 1700 - 5000 M$_\odot$. Clusters in these two groups are classified as {\it moderately rich} clusters. These groups have 122 clusters.\\
(4) Group V: These clusters have a large number of stars as members and are also massive, with mass $>$ 5000 M$_\odot$. Clusters in this group are classified as {\it rich} clusters. This group has only 9 clusters in our study. \

These classification and mass range are also presented in columns, 6 and 7 of Table \ref{number_criteria}. 
The mass range shown in the table re-confirms that the star clusters in the LMC have a very large mass range.  It can be seen that the mass range we have chosen to group clusters, is similar to the mass range shown in figure 3 of \citet{baum2013}. They restricted their analysis to clusters more massive than 5000 M$_\odot$, which are basically {\it rich} clusters according to our classification criteria. \citet{grijs2013} analysed different cluster mass range for estimating cluster mass functions in the LMC. In their table 2, they used three cluster groups to estimate the cluster mass function, which are, clusters with M$_c$ $>$ 1000, 3000 and 10000 M$_\odot$ respectively. The clusters in the first group are similar to very poor clusters, those in the second group are a combination of poor and moderate clusters, and the last group consists of rich clusters. Thus, the classification introduced in this study is in tune with the grouping done by earlier studies. 

There were previous attempts to group clusters into mass ranges, as mentioned in the above paragraph, primarily to study the slope of the cluster mass function. In these studies, the authors estimated the mass for individual clusters and then estimated the mass distribution. In this study, we classified each cluster according to the mass so that clusters can be preselected according to their mass to carry out studies of various properties of clusters which depend on its mass.  The clusters of various mass may vary in the formation mechanisms and hence the episodes for cluster formation across the mass spectrum may or may not be similar. Also, the survival of clusters is a function of mass and hence this classification will greatly help in understanding the dissolution of star clusters. Thus, this classification scheme will be very helpful to understand the formation, evolution as well as survival mechanism of these groups in the LMC. We also note that we have statistically significant number of clusters in these groups, such that we can study their properties as a function of group. We analyse the properties of these groups in the later sections and demonstrate the usefulness of classification. We plan to estimate the mass of individual clusters and study the mass function of each group, in the future.  

Here, we compare the mass range of clusters with that of open clusters in the Galaxy. \citet{pisku2008}, in their figure 2, show the distribution of mass of 650 open clusters with mass range 50 M$_\odot$ to 10$^5$ M$_\odot$. Their figure 5 shows the mass function of Galactic open clusters massive than $\sim$ 300 M$_\odot$. \citet{chou2015} found that the mass of the LMC clusters which were identified as probable asterisms were $\sim$ 300 M$_\odot$.
\citet{lamers2005}, in their figure 9 show the mass distribution of clusters in the solar neighbourhood, which is similar to the mass range in Table 1. Thus, the LMC cluster mass range is very much similar to the mass range of the Galactic open clusters, at least in the low mass end of the cluster mass distribution.

\section{Error estimation}
In this study, we have used a quantitative method to estimate age and reddening and hence we can clearly estimate the error in the estimation of these parameters.  The error associated with the estimated age depend upon the photometric errors, errors in estimating extinction, binning resolution along the magnitude as well as colour axes. We computed error in the estimation of extinction and age using the method of propagation of errors. Error in the estimation of extinction and age are given by the following relations:\\
$\sigma A_V$ = $ 2.48  \sqrt {\sigma{(V-I)}^2 + {(V-I)}_{bin}^2}$ \\
$\sigma$$M_V$ = $ \sqrt {\sigma V^2 + V_{bin}^2 + \sigma A_V^2}$ \\
$\sigma$age = constant x $\sigma$$M_V$ \\
where, $V_{bin}$ \& ${(V-I)}_{bin}$ are half the bin sizes used for magnitude \& color binning,  $\sigma$$M_V$ is the error in absolute magnitude, $\sigma$$A_V$ is error in the estimated extinction and $\sigma$age is the error in estimated age in a logarithmic scale (log(t)). This analyses yields that the the maximum error in age is 0.2. We have adopted this value as the error in the estimation of age. As the reddening is estimated from the upper MS, the photometric error could be considered negligible. Thus the error in the estimation of reddening, E(V$-$I) is taken as the bin size, 0.1 mag.\\

\section{Results and Discussion}

\begin{figure}
\includegraphics[width=\columnwidth]{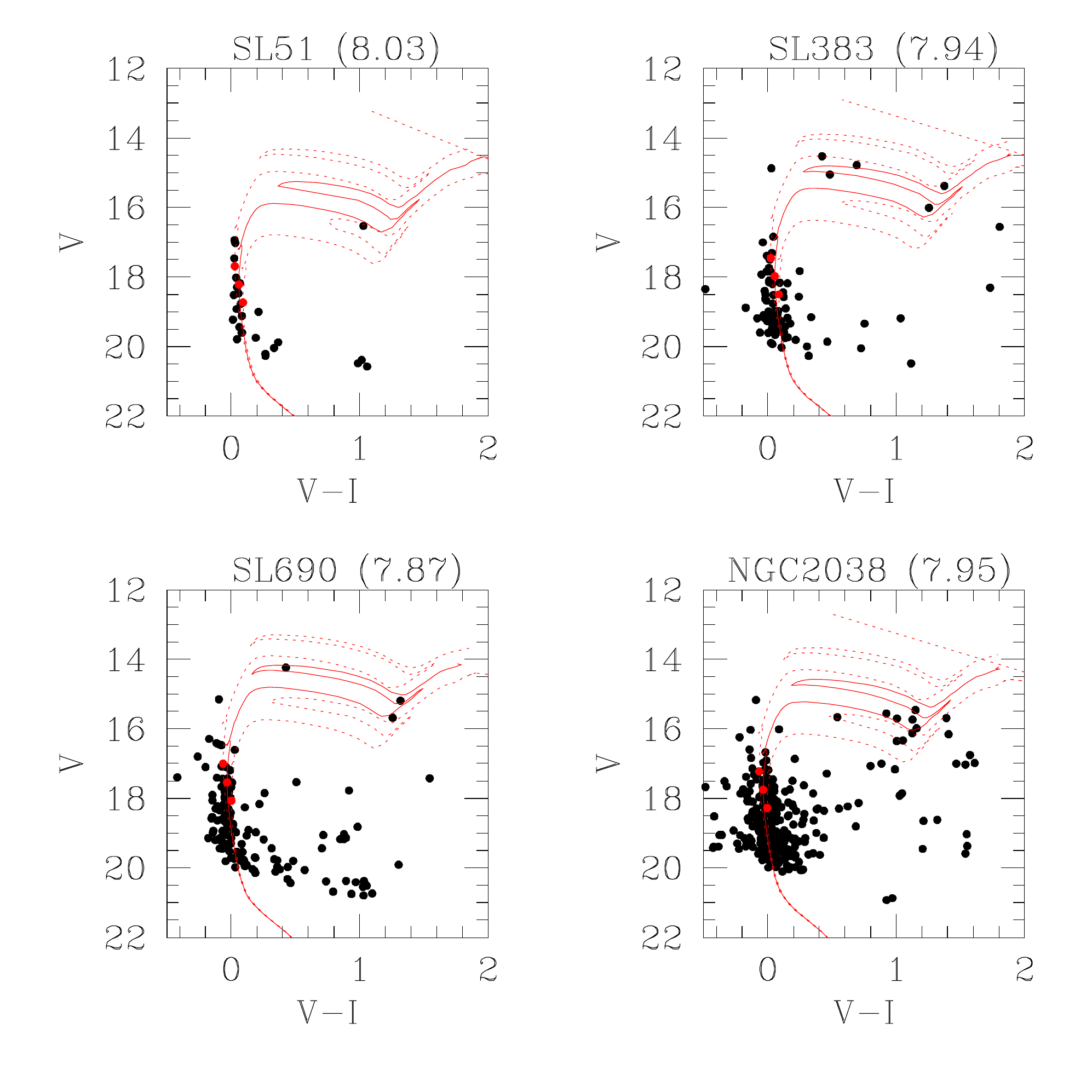}
\caption{An example of cleaned cluster CMD from four groups with over plotted isochrones of estimated ages (red solid line) along with error (red dotted line). The top left panel shows the CMD of SL 51 from very poor group, top right panel shows the CMD of SL 383 from poor group, bottom left panel shows the CMD of SL 690 from moderate group and bottom right panel shows the CMD of NGC 2038 from rich group. CMDs of all 1072 clusters are available online.}
\label{cmds_four_grp}
\end{figure}

In this study, we have classified and estimated the age and reddening of 1072 star clusters using a semi-automated quantitative method. 
Although there have been many studies of the LMC star clusters in the past this is the first attempt to classify LMC star clusters into groups based on richness/mass and a quantitative method is introduced for its parameterisation. 
The results of this study are presented as catalog of clusters, will be available as an on-line catalog. A sample of this catalog is shown in Table 2. This catalog contains the name of the cluster, central coordinates (RA and Dec) as in B08, size of the cluster taken from B08,  estimated reddening and age, earlier estimations of ages by G10, PU00, \citet{palma2016_catalog}, \citet{piatti2014_vmcXII,piatti2015_vmcXVI} and the designated group number (I-V) repectively. We also present the CMDs of all the 1072 clusters in the same order as in the Table 2. The cluster CMDs are field subtracted, over plotted with the M08 isochrone, with an uncertainty of log(t) = $\pm$0.2 for the estimated age and reddening. The location of the turn-off as per the isochrone is shown as red dot, on the MS in the CMD. In some of the CMDs, data points are shown in different point types. These are the clusters for which we identified the cluster sequence after performing field star subtraction with three different annular field regions. Most probable cluster members are those which survive at least two separate field subtraction process and these can be identified as those with at least two point types appearing one on top of another. These are also available in the online version. We believe that the catalogue as well as the CMDs presented here will be a valuable resource for various studies involving star clusters in the LMC, such as understanding the hierarchical formation of clusters in selected regions of the LMC. As an example, we present one CMD each from the four groups of clusters in Figure \ref{cmds_four_grp}. The top left panel shows the cluster SL 51, which belongs to the very poor cluster. The top right panel shows the CMD of SL 383, which is a poor cluster. The bottom left panel shows the CMD of the moderate cluster SL 690, whereas the bottom right panel shows the CMD of a rich cluster, NGC 2038.

We estimated parameters of 308 clusters for the first time. Among these new estimates, 156 are group I (very poor) clusters, 111 are Group II (poor) clusters, 39 belong to groups III \& IV (moderate) clusters and 2 clusters belong to group V (rich).  
In the Table 2, these clusters can be identified as those with columns 7-10 as blank.

\subsection{Comparison with previous estimations}
\begin{figure}
\includegraphics[width = \columnwidth]{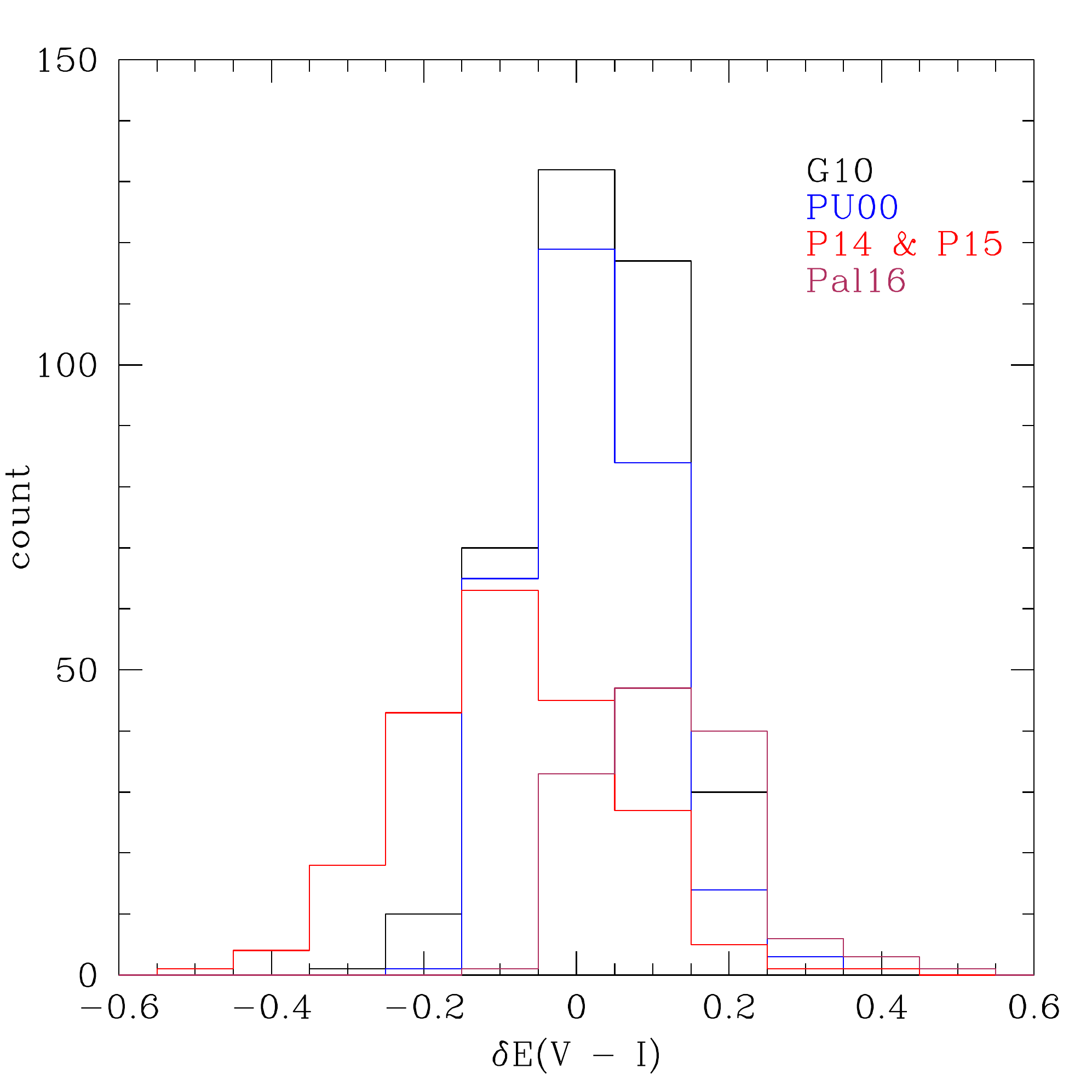} 
\caption{The distribution of difference in reddening (this study $-$ previous estimation). The previous studies used for comparison are that of G10 (black),PU00 (blue), \citet{piatti2014_vmcXII,piatti2015_vmcXVI} (red) (in the figure mention as P14 $\&$ P15)  and \citet{palma2016_catalog} (maroon) (in the figure mentioned as Pal16).}
\label{compare_previous_study}
\end{figure}

Among the clusters studied here, many clusters are common with various previous studies. There are 366 clusters in common with G10, 287 clusters with PU00, 208 clusters with \citet{piatti2014_vmcXII,piatti2015_vmcXVI} and 131 clusters with \citet{palma2016_catalog}. 
We compared our estimations of reddening and age with these studies. The Figure \ref{compare_previous_study} shows the difference in reddening where the difference is estimated as (our study - previous study). Our estimates compare very well with the reddening estimates of previous estimations, except in the case of \citet{piatti2014_vmcXII,piatti2015_vmcXVI}. The difference in reddening with the estimation of G10 and PU00 are centered around zero, the difference with respect to \citet{palma2016_catalog} is centered at 0.1 mag, which is within the error (1 $\sigma$) in our estimation. The difference in reddening with respect to the estimation of  \citet{piatti2014_vmcXII,piatti2015_vmcXVI} is centered around $-$0.1 mag, which is 1 $\sigma$ error in our estimation, suggesting that the reddening estimated by \citet{piatti2014_vmcXII,piatti2015_vmcXVI} is slightly higher than our estimates. We also note that the study by \citet{piatti2014_vmcXII,piatti2015_vmcXVI} is based on near-IR data, whereas the rest of the studies are based on optical data. 

\begin{figure}
    \centering
    \begin{subfigure}[t]{0.49\textwidth}
        \centering
        \includegraphics[width=\columnwidth]{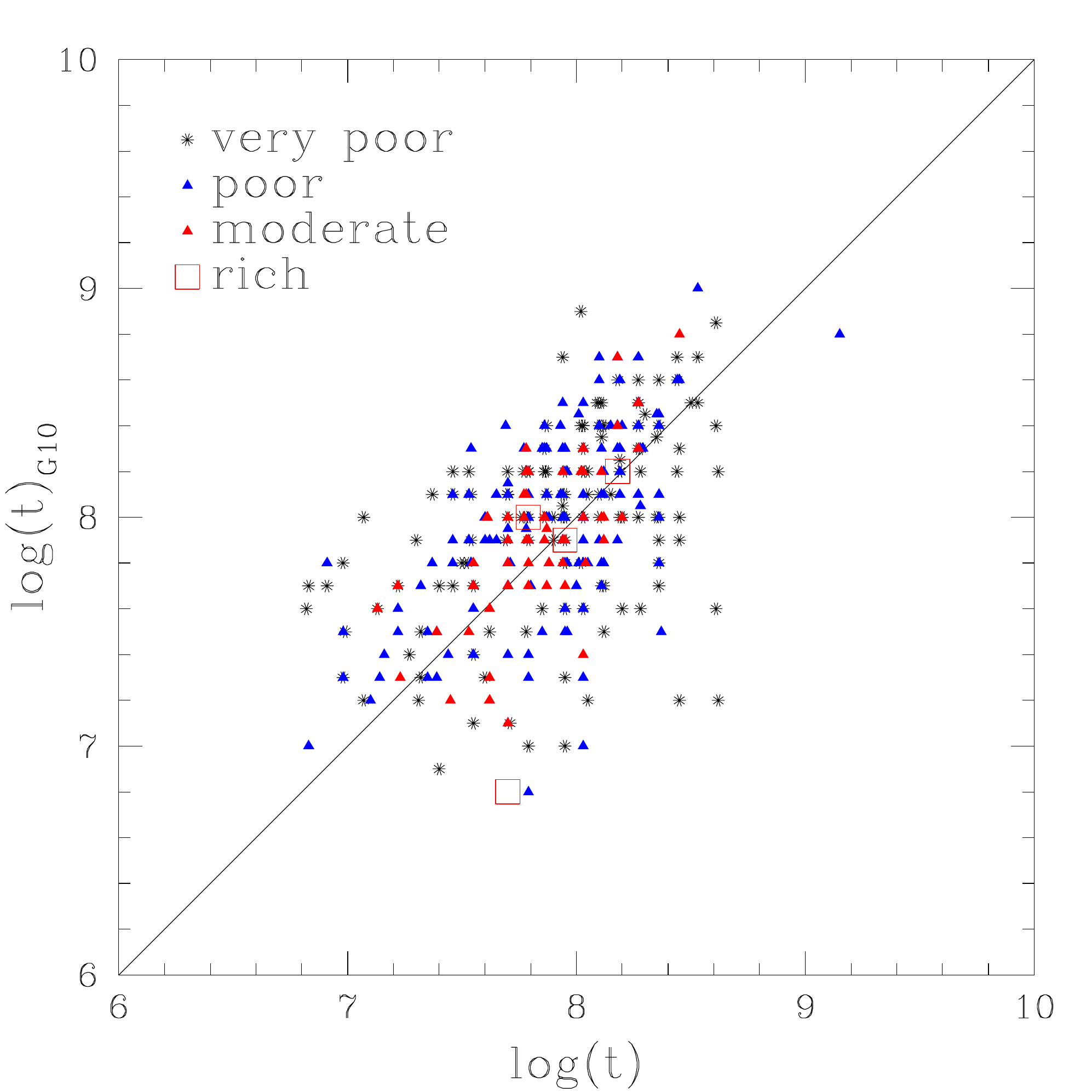}
        \caption{}
    \end{subfigure}%
    ~ 
    \begin{subfigure}[t]{0.49\textwidth}
        \centering
        \includegraphics[width=\columnwidth]{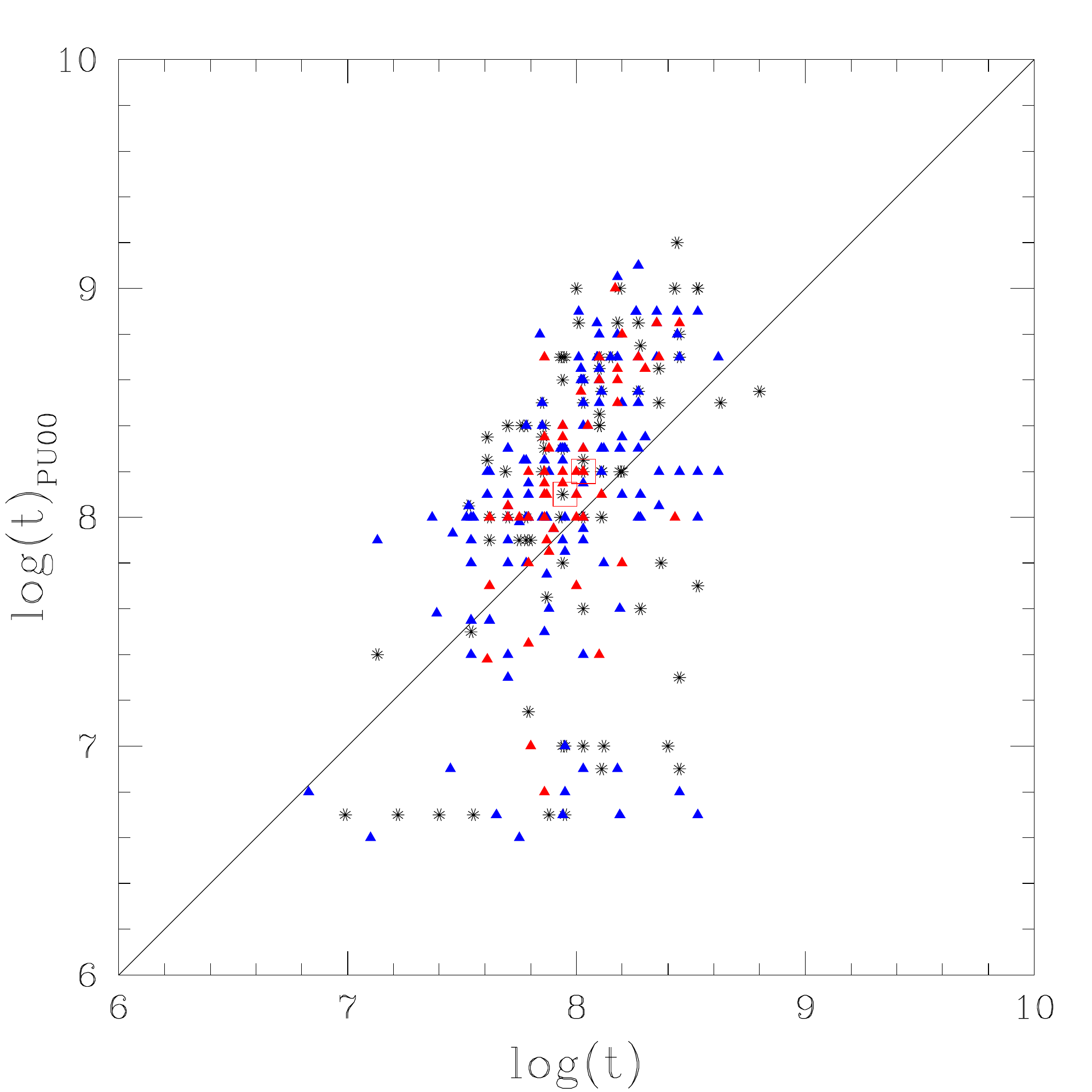}
        \caption{}
    \end{subfigure}
    
    \begin{subfigure}[t]{0.49\textwidth}
        \centering
        \includegraphics[width=\columnwidth]{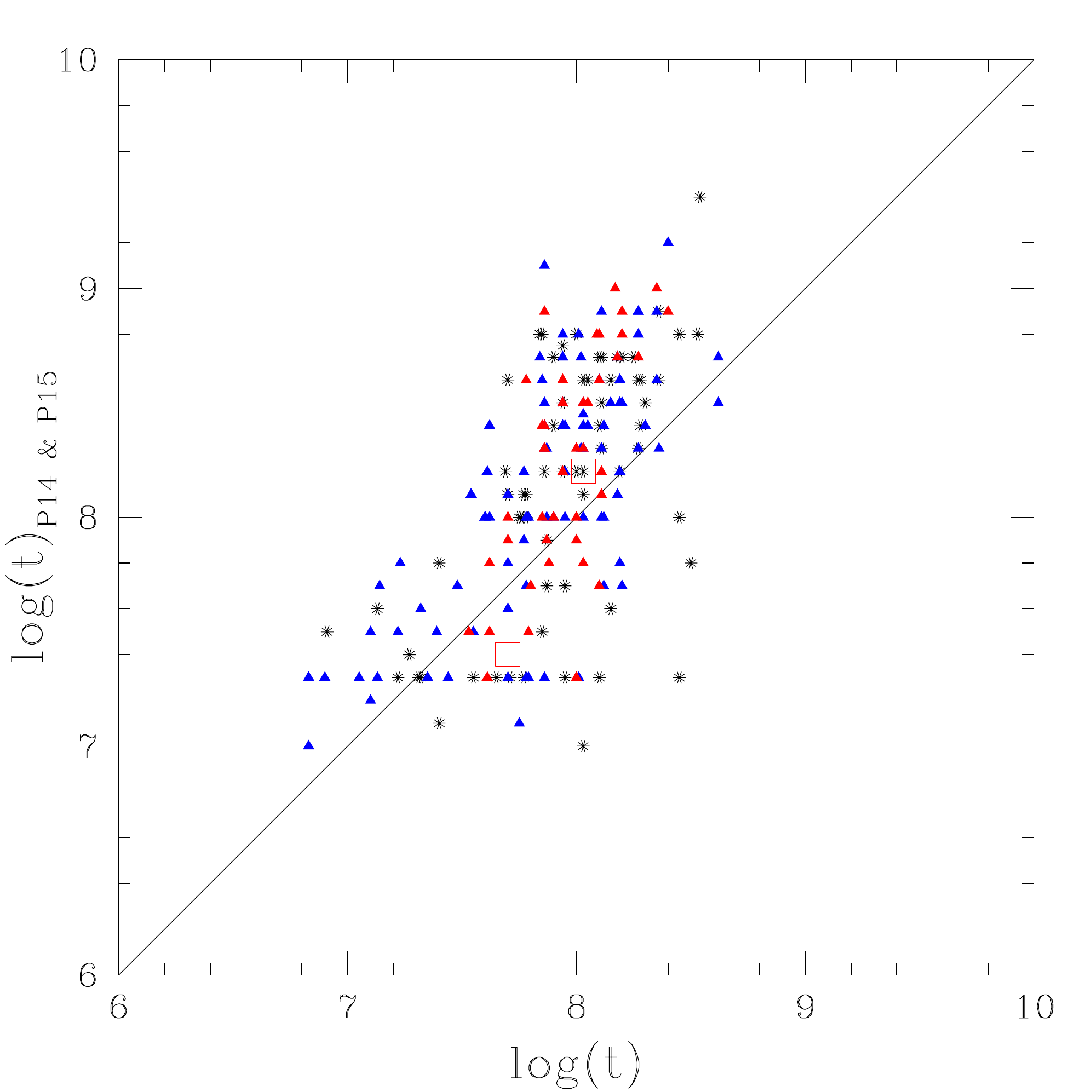}
        \caption{}
    \end{subfigure}%
    ~ 
    \begin{subfigure}[t]{0.49\textwidth}
        \centering
        \includegraphics[width=\columnwidth]{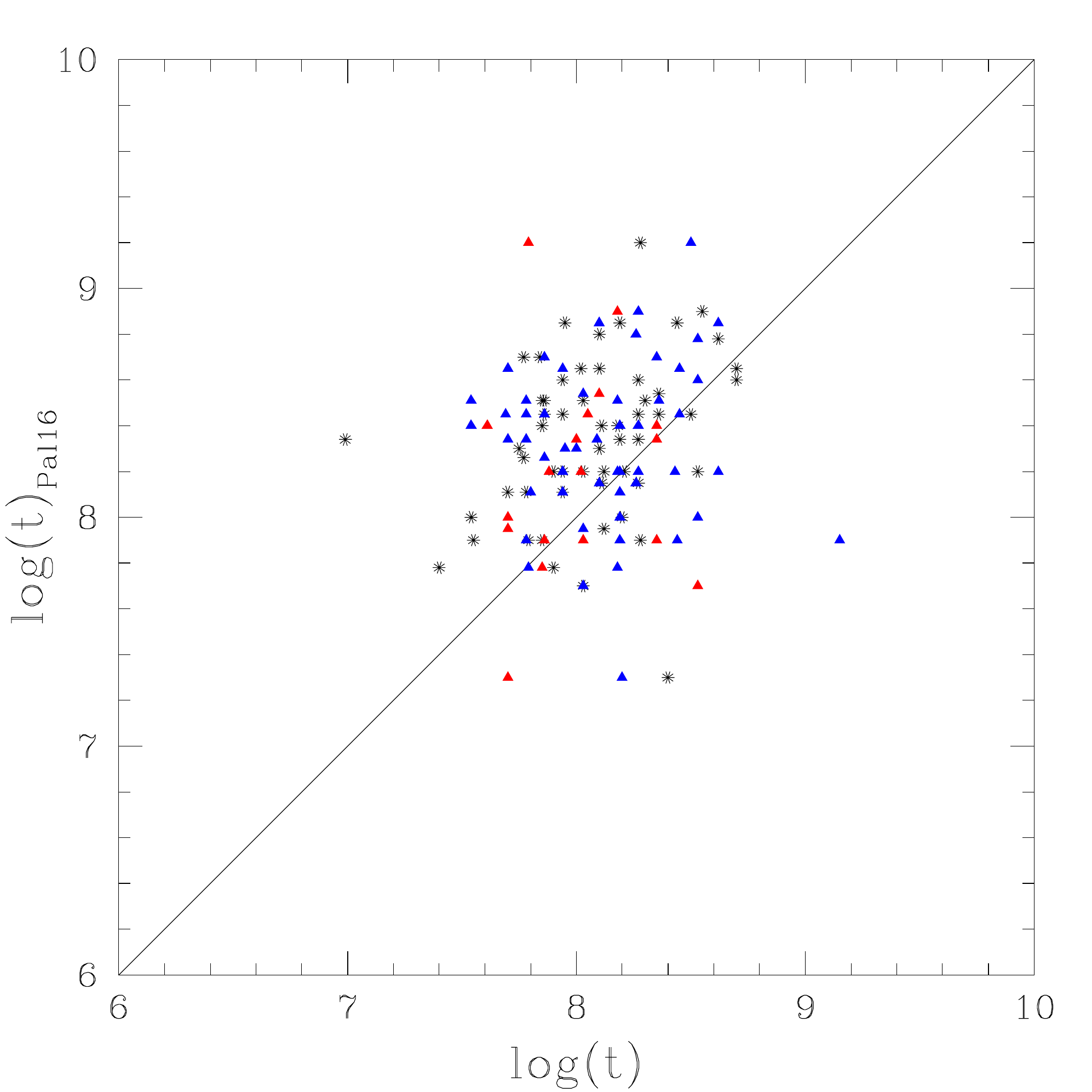}
        \caption{}
    \end{subfigure}
    \caption{The plots show the comparison of estimated age with previous age estimation. The previous studies used to compare our age estimation are that of G10 (Fig.a),PU00 (Fig.b), \citet{piatti2014_vmcXII,piatti2015_vmcXVI} (Fig.c) and \citet{palma2016_catalog} (Fig.4). Different point types are used for different group of clusters as mention in the Fig.a.}
\label{our_age_vs_others}
\end{figure}

We compared our age estimations with the above mentioned 4 estimates. In Figure \ref{our_age_vs_others}, we have shown our estimation in the x-axis and the previous study in the y-axis, such that comparison with the 4 studies are shown in 4 different panels. A straight line with slope equal to one, is also shown indicating the location of clusters with identical age estimation in both the studies. We have also shown clusters in different groups using different symbols as explained in the Figure \ref{our_age_vs_others} caption. It can be seen that our age estimations match very well with the estimation of G10, except in a few cases (top left panel (a)). The observed scatter is found to be relatively large for the very poor clusters which is found to decrease for the poor and moderate clusters. 
In the top right panel, we compared our estimations with those of PU00. In this plot, we can detect two patterns, one almost along the straight line and a horizontal one. The horizontal pattern consists mostly of very poor and poor clusters and two moderate clusters. This pattern suggests that we estimated a range of ages for these clusters, whereas PU00 estimated constant age about 10Myr. We visually checked all the clusters in this pattern and re-confirmed the estimated parameters. We also notice that the PU00 estimated relatively older ages for clusters older than 100 Myr, whereas the clusters younger than 100 Myr have ages within the error (as seen from the pattern close to the straight line). We would like to point out that the PU00 used OGLE II data which has lesser resolution than the OGLE III data. As all these clusters are located in the bar region of the LMC, effect of crowding may be more in the OGLE II data. Also authers used isochrone model by Bertelli et al. (1994). We speculate that the difference in age estimation noticed here could be due to these reasons. 
In the panel (c), we compare our estimation with those of \citet{piatti2014_vmcXII,piatti2015_vmcXVI}. There is a large range in age and most of the common clusters are in the poor and moderate category. For clusters younger than 100 Myr, the ages compare very well. For older clusters, we estimated younger ages relative to the estimation of \citet{piatti2014_vmcXII,piatti2015_vmcXVI}. 
 In the panel (d), we compare our estimation with \citet{palma2016_catalog}. Thus clusters are mostly older than 100 Myr and we do not detect any systematic trend, but a large scatter. 

 In summary, the comparison suggests that the age and reddening estimations compare well with the previous estimations, though we do detect discrepancies in the case of a few clusters. The average difference in the estimation of age with G10 is zero whereas age difference with other literature peaks at 0.2. In comparison with PU00 we get maximum dispersion in age estimation of 2.0 in log scale whereas with other studies maximum dispersion is 1.0.
The differences could be due to various reasons, such as difference in the data used, difference in the isochrones and the assumed metallicity and the adopted field star removal process. For example, \citet{palma2016_catalog} and \citet{piatti2014_vmcXII,piatti2015_vmcXVI} used data
from bigger telescopes and also different photometric system. \citet{chou2015} found that the brighter stars were saturated in the data (which is similar to the data used by \citet{palma2016_catalog}) and therefore estimated a relatively older age for young clusters. On the other hand, \citet{piatti2014_vmcXII,piatti2015_vmcXVI} used near-IR data from the VMC survey and hence a direct comparison is difficult. \citet{piatti2014_vmcXII,piatti2015_vmcXVI} used the metallicity Z=0.006 and isochrones from Bressan et al. (2012). They also mention that the ages are not sensitive to the adopted metallicity. There are also differences in the isochrones used. \citet{palma2016_catalog} used theoretical isochrones computed for the Washington system by the Padova group (Girardi et al. 2002; Bressan et al. 2012) and Geneva group (Lejeune \& Schaerer 2001). They have also shown in their figure 5 that age estimation does not differ much by these two isochrone model. \citet*{glatt2010} had also used two different isochrone models $:$ Padova isochrones (Girardi et al. 1995) and Geneva isochrones (Lejeune \& Schaerer 2001) in their study.

\subsection{Reddening distribution across groups}
\begin{figure}
\includegraphics[width=\columnwidth]{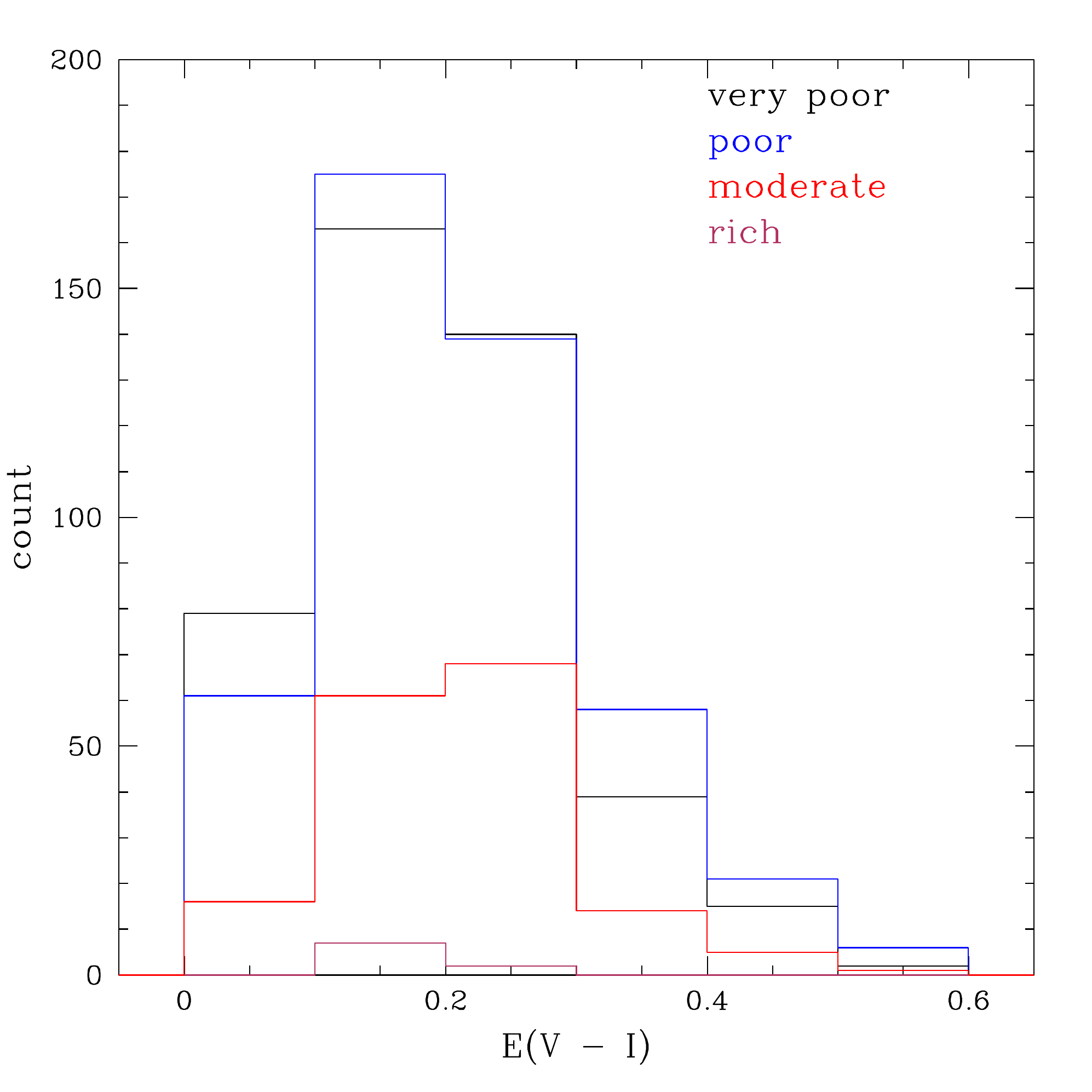}
\caption{Histogram of estimated reddening for very poor (black), poor (blue), moderate (red) and rich (maroon) clusters. For all the four groups reddening peaks around 0.2 magnitude. }
\label{reddening_histogram_diff_group}
\end{figure}

In this section, we compare the reddening of clusters across various groups. 
The distribution of the estimated reddening is shown in Figure \ref{reddening_histogram_diff_group} for various cluster groups. The estimated reddening is found to be in the range 0.05 - 0.55 mag. The distributions have a peak between the values 0.1 - 0.3 mag in E(V$-$I) for all the four groups. The distribution is also found to be similar for all the 4 groups of clusters. We also notice that the tail of the distribution towards higher reddening is also similar for the very poor, poor and moderate clusters.

\begin{figure}
\includegraphics[scale=0.35]{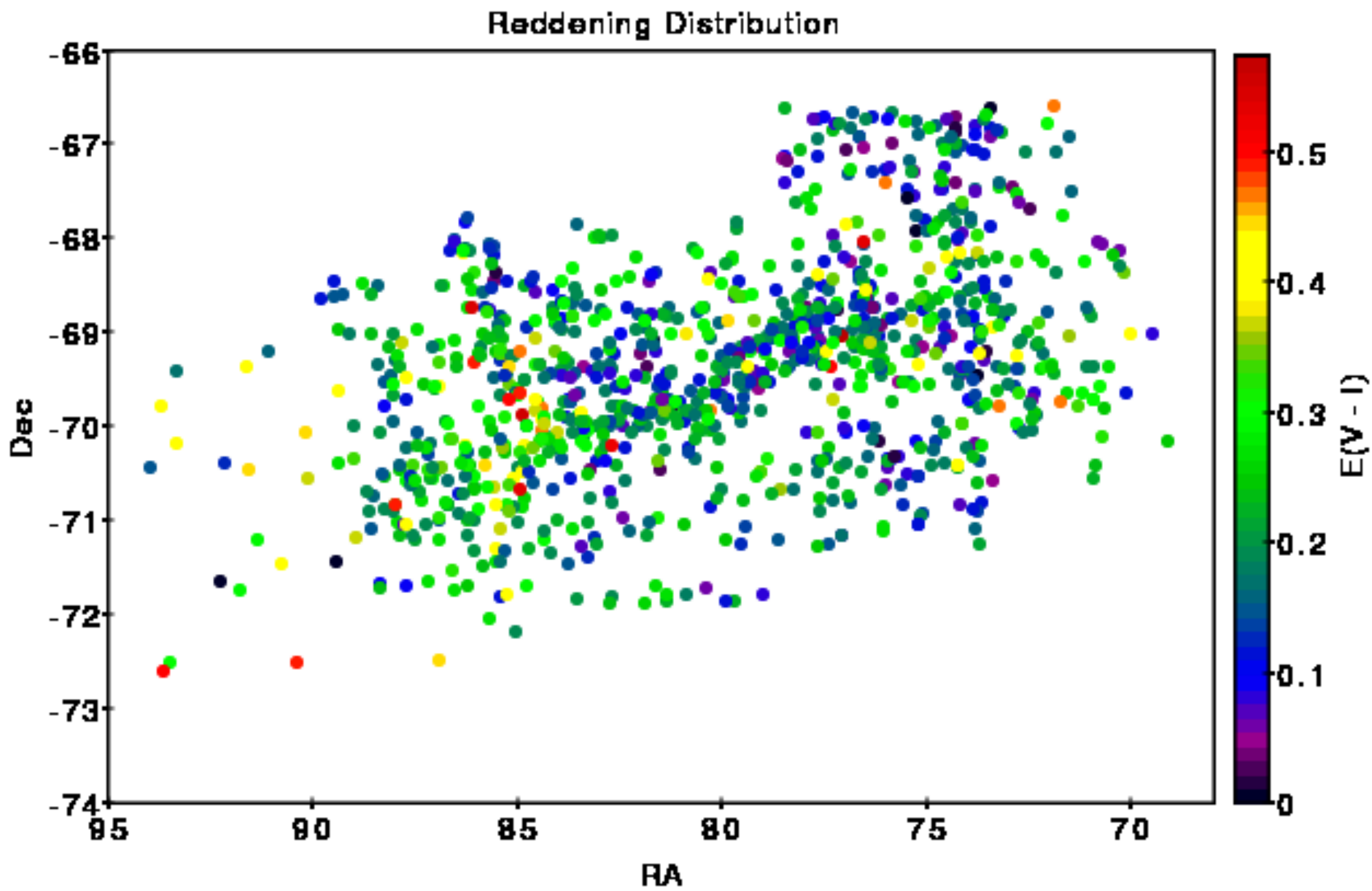}
\caption{The spatial variation of estimated reddening for all the clusters.}
\label{reddening_dist}
\end{figure}

 The distribution of reddening as a function of position is shown in Figure \ref{reddening_dist}. Here we have shown all the groups together, as we did not find any difference in reddening among the groups. The figure shows that the reddened clusters are, in general, located in the central regions. The bar region has a larger range in reddening. The clusters in the south have relatively less reddening, whereas a few clusters located in the south-east corner and east have relatively large reddening. Most of the clusters located near the center of the bar show relatively less reddening, when compared to those located near the ends of the bar. We compared the estimated reddening with those estimated for the field regions by \citet{indu2011}, which is a high resolution
map covering the central LMC. We did not find any significant variation in reddening with respect to the field reddening.

\subsection{Age distribution across groups}

\begin{figure}
\includegraphics[scale=0.59]{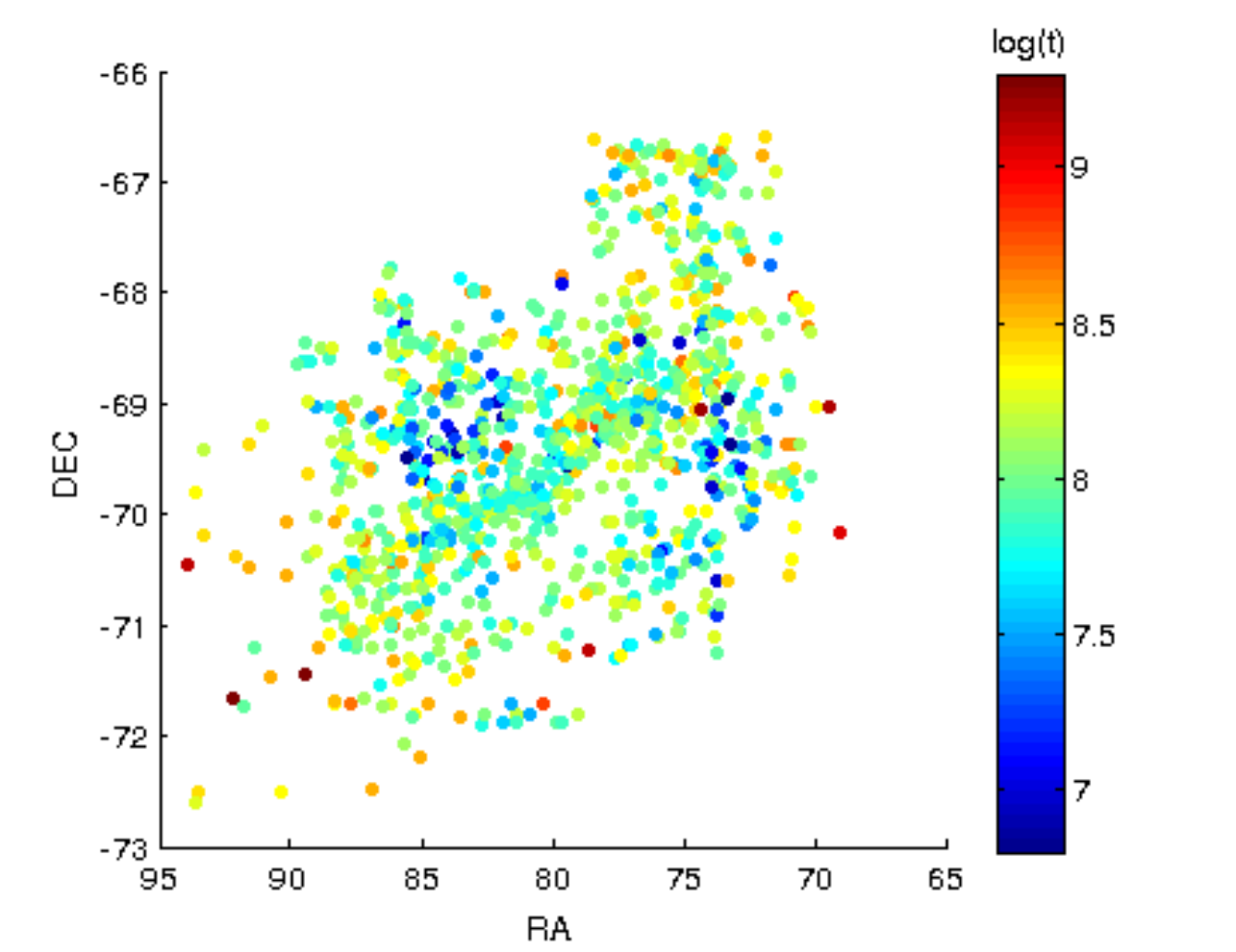}
\caption{Plot shows spatial distribution of estimated ages of all the 1072 clusters.}
\label{age_dist_total}
\end{figure}

\begin{figure}
\includegraphics[width=\columnwidth]{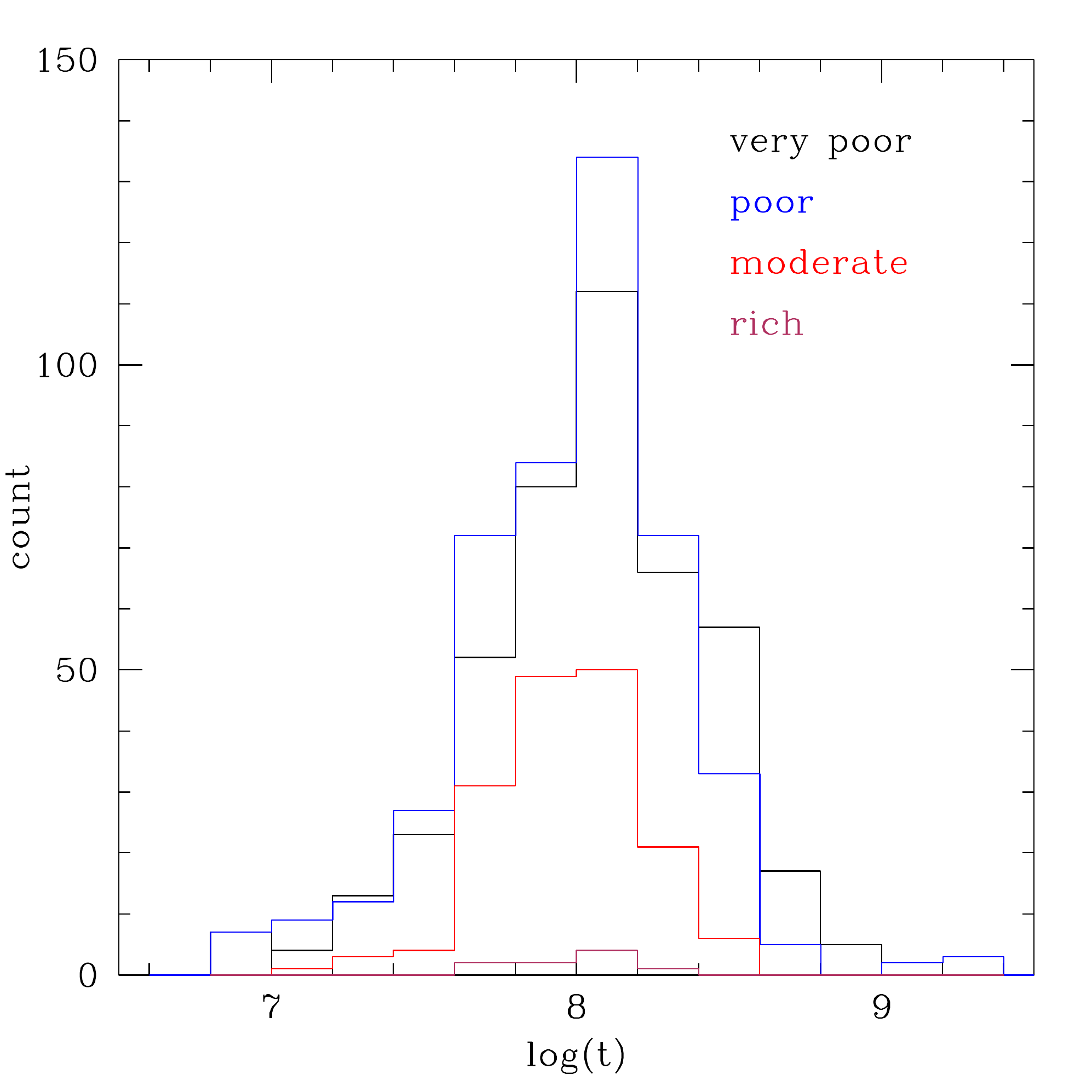}
\caption{Distribution of cluster age. Age distribution of very poor, poor and clusters peak between log(t) = 8.0 - 8.2 where moderate clusters has a broader peak (log(t) = 7.8 - 8.2).}
\label{age_histogram}
\end{figure}

The spatial distribution of age of all clusters studied here, is shown in Figure \ref{age_dist_total}. The plot indicates the recent sites of star formation in blue and the older clusters are shown as brown. The younger clusters are found to be located in the inner part of the LMC, whereas the older clusters show a more spread out distribution. We can clearly see the southern arm like pattern with the presence of some young clusters. 
As we have a large sample of clusters, we can study the age distribution of the clusters studied here, as a function of groups. The distribution of the estimated age is shown in Figure \ref{age_histogram}. The age distribution of various groups are also shown separately in the figure. The figure shows that the peak of the cluster age for the poor and the very poor clusters lie in the bin log(t) = 8.0 - 8.2. This peak in the age distribution can be considered to be at 125 $\pm$ 25 Myr. We have binned the distribution with a bin size of 0.2 in log scale, as this is the same as that of the error in age estimation. This peak is similar to those found by G10 and PU00. The moderate clusters show a broad peak between log (t)= 7.8 - 8.2 with the mean at 100 Myr. The rich clusters also show a peak similar to the poor and very poor clusters, though their number is very small. This peak is attributed to the interaction between the LMC and the SMC, about 200 - 300 Myr ago \citep{besla2012}.

The point to be noted is that on the whole, the age distributions of poor and very poor clusters are very similar. A careful inspection reveals that the very poor cluster group has more number of older clusters (log(t) = 8.4 - 8.8), whereas the poor cluster group has more of younger clusters (log(t) = 7.6 -   7.8). This trend
is found to be more pronounced in the age distribution of moderate clusters, where the peak itself extends to younger ages. The above mentioned trend is observed only up to log(t) = 7.6. Relatively less number of moderately rich clusters are found in the bins older than log (t) = 8.0, up to log (t) = 8.6. If we consider a progressive formation of clusters 
from the age of log(t) = 8.6 to younger ages, the very poor clusters are formed first, followed by the formation of poor
clusters and then the moderate clusters. Thus, in the inner LMC, our study suggests a hierarchical formation of clusters from low to high, in terms of mass of the clusters. This suggestion needs to be tested in more detail to validate it.

\begin{figure}
    \centering
    \begin{subfigure}[t]{0.49\textwidth}
        \centering
        \includegraphics[width=\columnwidth]{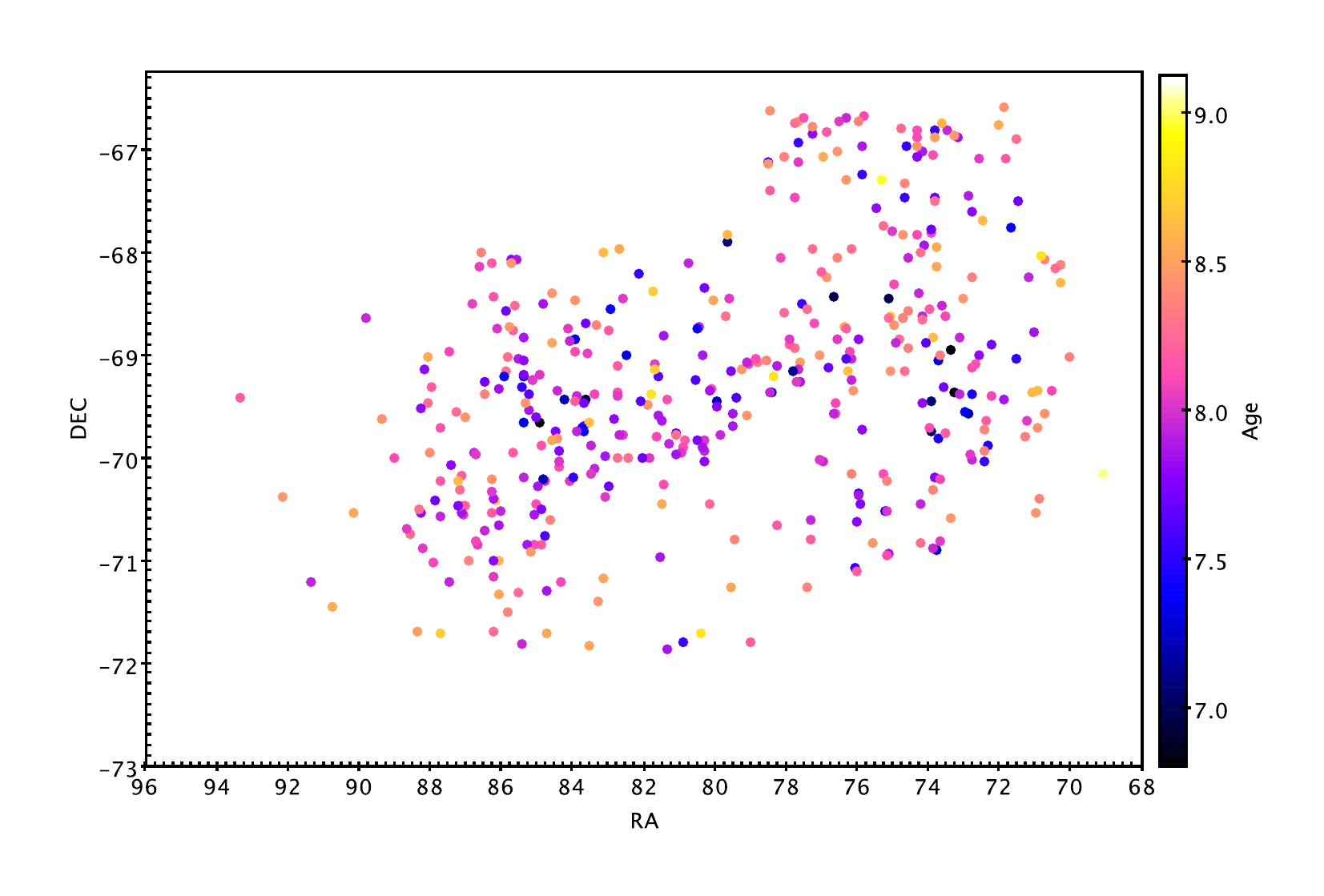}
        \caption{}
    \end{subfigure}%
    ~ 
    \begin{subfigure}[t]{0.49\textwidth}
        \centering
        \includegraphics[width=\columnwidth]{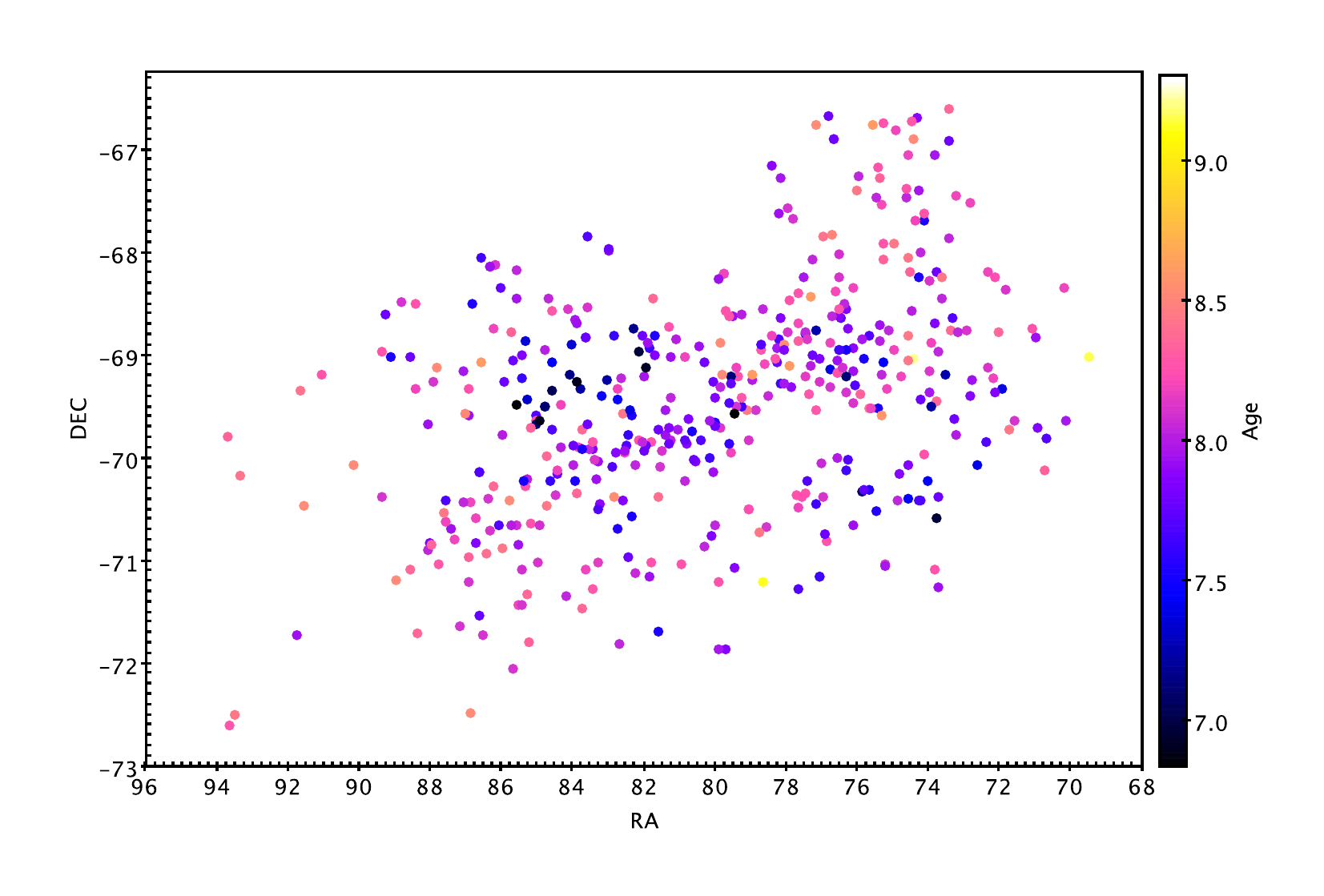}
        \caption{}
    \end{subfigure}
    
    \begin{subfigure}[t]{0.49\textwidth}
        \centering
        \includegraphics[width=\columnwidth]{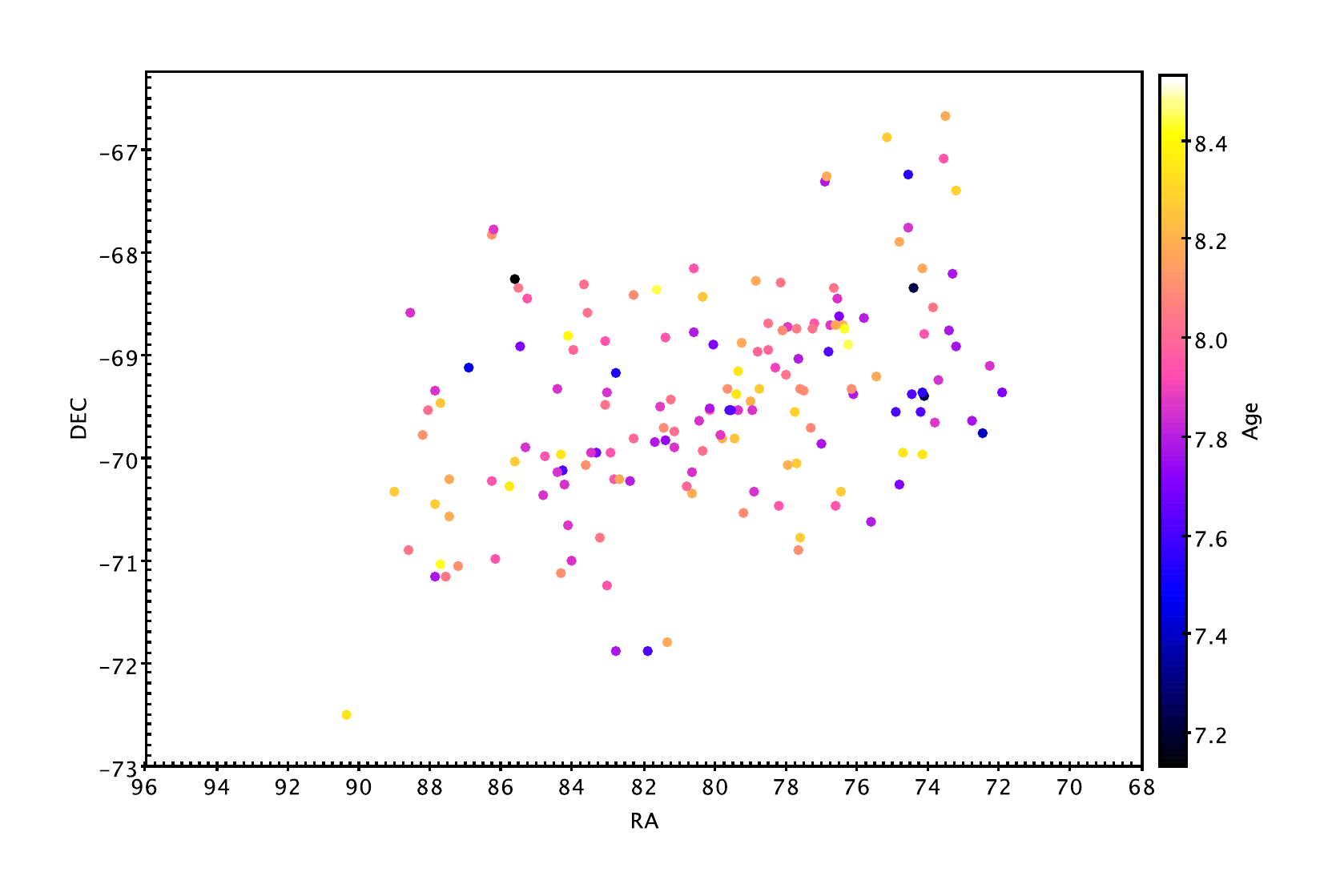}
        \caption{}
    \end{subfigure}%
    ~ 
    \begin{subfigure}[t]{0.49\textwidth}
        \centering
        \includegraphics[width=\columnwidth]{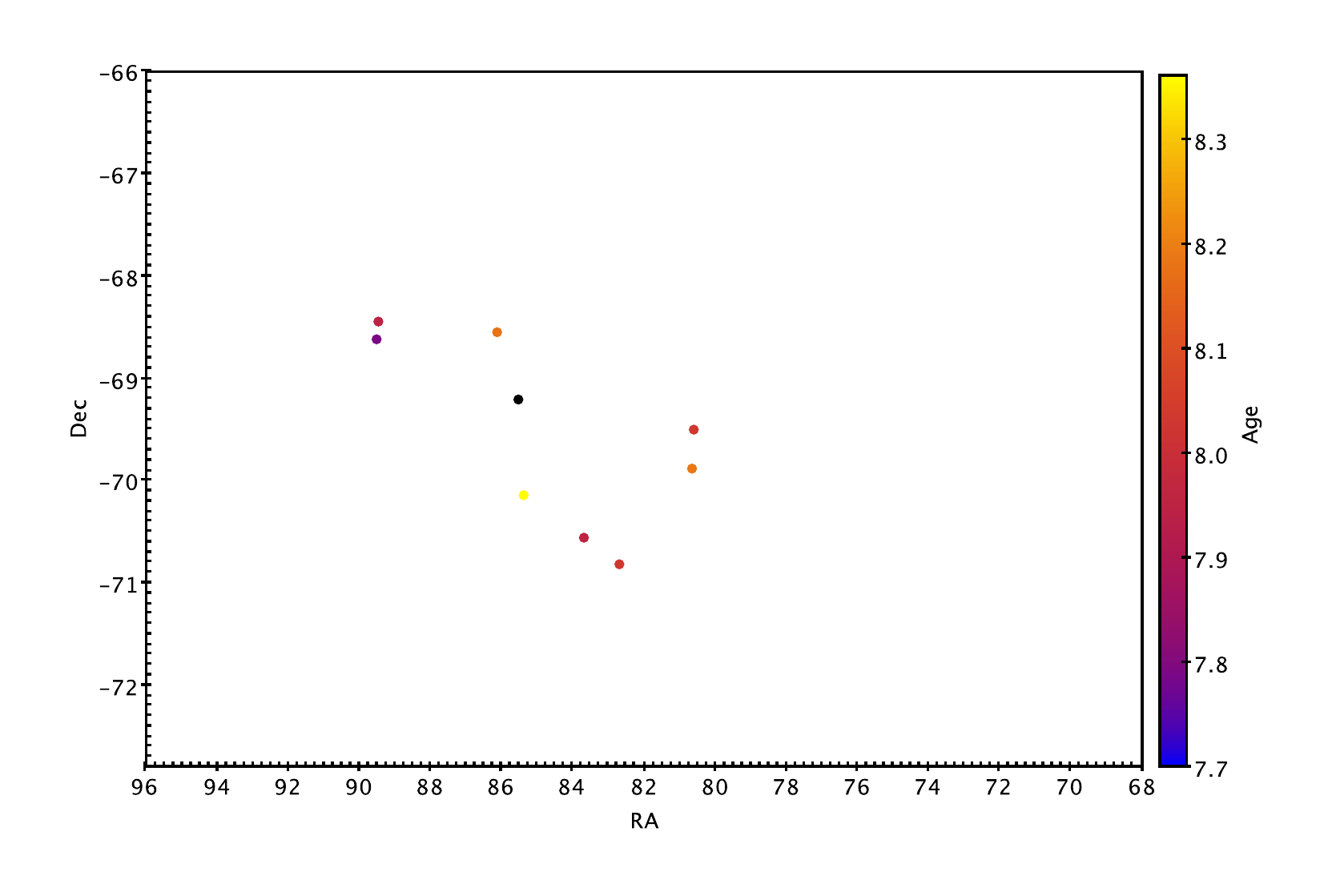}
        \caption{}
    \end{subfigure}
    \caption{Spatial distribution of ages in the LMC field for (a) very poor, (b) poor, (c) moderate and (d) rich clusters.}
\label{spatial_age_dist_diff_grp}
\end{figure}

In order to understand the spatial distribution of clusters as a function of age and richness, we have shown the spatial distribution of cluster ages separately for four richness groups in the Figure \ref{spatial_age_dist_diff_grp}. The left top panel shows the spatial distribution of very poor clusters where the color coding is according to the age.  
It can be seen that the very poor clusters are formed all over the inner LMC. We cannot see the bar clearly, suggesting that the very poor clusters are not preferentially formed in the bar. The top right panel shows the distribution of poor clusters. Their distribution not only delineates the bar clearly, but also are found located along the bar in the inner LMC. The bottom left panel shows the location of moderate clusters. They are preferentially 
located in the inner LMC and in the bar region. The rich clusters are shown in the bottom right panel. Thus, we see that the more massive clusters, are preferentially formed in the inner regions and along the bar, whereas the low mass clusters belonging to the very poor cluster group are found in a relatively scattered way. Thus we demonstrate that the formation
of clusters in the inner LMC differ as a function of total mass of the cluster. We have been able to demonstrate it because of the classification scheme adopted in this study. It will be interesting to study the details of cluster formation as a function of mass in the regions outside the area of this study.

\onecolumn

\twocolumn

%
%
%

\onecolumn

\twocolumn

\begin{figure}
\includegraphics[width=\columnwidth]{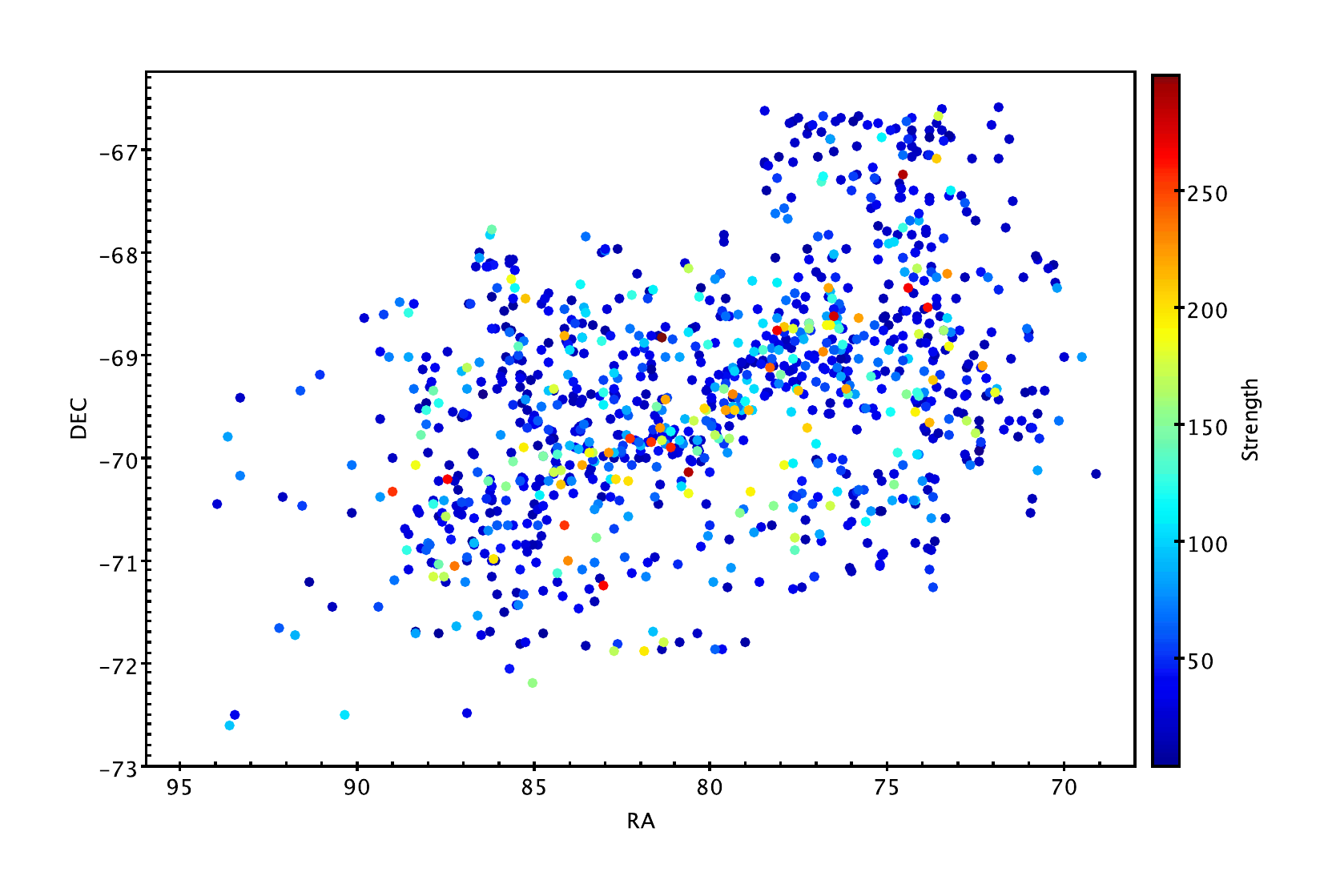}
\caption{Distribution of all clusters as a function of strength in the LMC field.}
\label{strength_dist}
\end{figure}

In order to elaborate this point, we have shown the spatial distribution of all clusters color coded according to number of members in the clusters in Figure \ref{strength_dist}. We can see that the clusters with members less than $\sim$ 50 are spread across a larger area, whereas those with more number of stars are found in the bar and inner regions. In this figure, we also find that the bar region is delineated very well suggesting that, in the inner regions, the clusters are preferentially formed in the bar of the LMC. The clusters are also found to form at the western end of the bar, in a region extending from north to south, roughly perpendicular to the bar. In the eastern side end of the bar, we identify two locations of cluster formation. Both the regions start from the bar and extend to the northern regions. The eastern most region has relatively less number of clusters, whereas the inner one has relatively large number of clusters. The inner one is coincident with the 30 Dor star forming region and the nearby super giant shells. We also notice an arm like feature in the south, populated with relatively low mass clusters. We have demonstrated that the cluster formation in the inner LMC has taken place in specific locations, in particular, the bar region of the LMC. We also demonstrated that the relatively rich/massive clusters are preferentially formed in the inner LMC, including the bar. These give valuable clues to the formation mechanism of clusters in the inner LMC, as a function of cluster mass.  

In the top panels of Figure \ref{spatial_age_dist_diff_grp}, we find that the younger clusters are preferentially formed in the inner regions, whereas the outer regions have older clusters. This trend is found in all the four richness groups of clusters. This means that, irrespective of the richness of a cluster, we identify a quenching of cluster formation form outside to inside. Combining the above two facts, we find a trend of cluster 
formation shifting to the central regions for younger ages and to massive clusters. In order to bring out this point clearly, we have created two videos (available online only), where the formation sequence of clusters are shown. In video-1, we show the sequence from younger to older ages, and in the video-2, we show the sequence from older to younger ages. The videos show clusters of the four groups in four different colours. These clearly suggest that the cluster formation has shrunk to the inner LMC. This is similar to that found for the field stars by \citet{indu2011}.
 
This section clearly brings out the advantage of classification of clusters based on groups. We have demonstrated that we are able to detect a significant difference in the spatial distribution of clusters as a function of mass. This study also suggests the importance of including the low mass clusters such as poor and very poor clusters in the cluster formation history of the LMC. The catalog as well as the classification can be used to understand the hierarchical formation of clusters in selected regions of the LMC.
 
\subsection{Cluster formation in the LMC bar}
\begin{figure}
\includegraphics[width=\columnwidth]{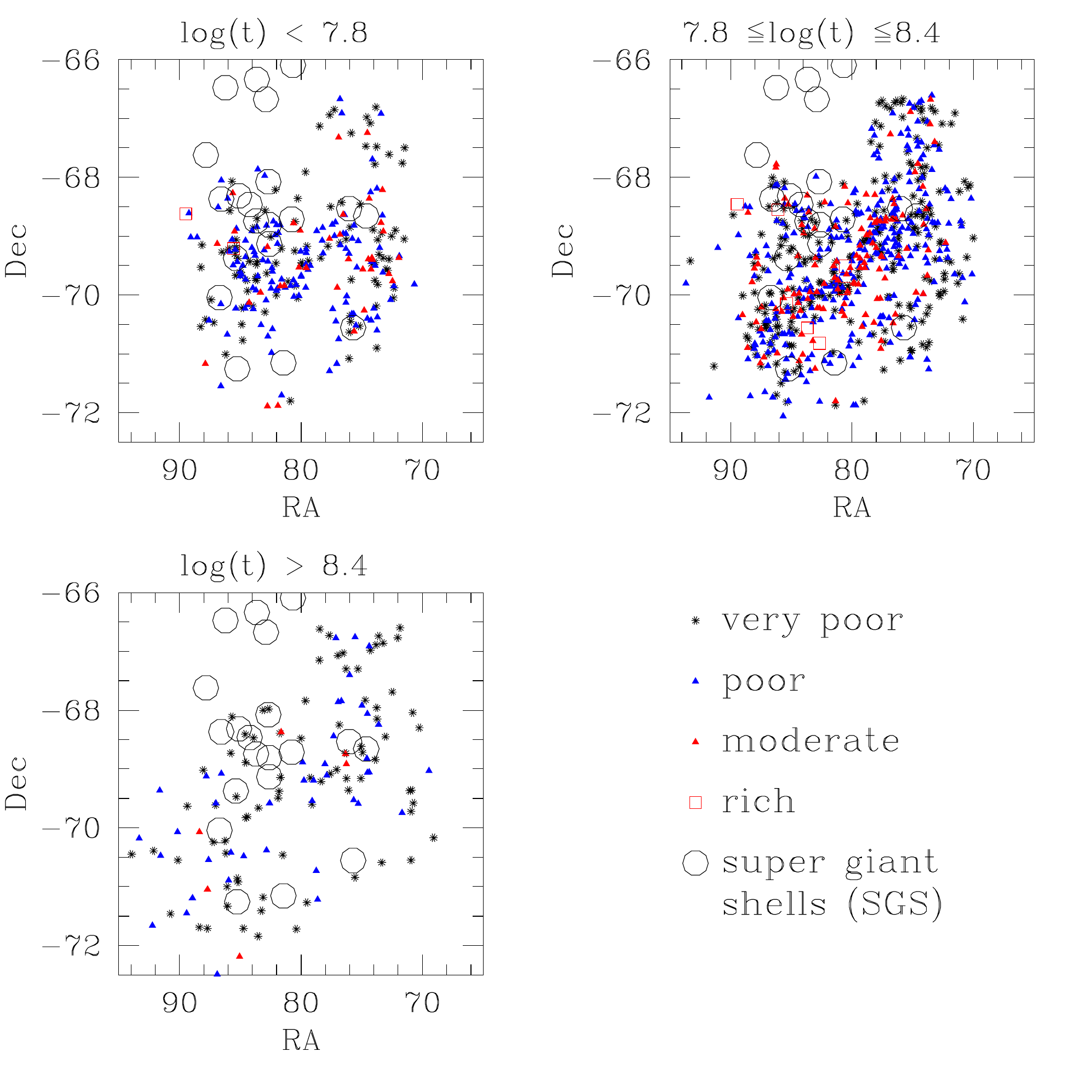}
\caption{Distribution of different group of clusters younger than 63 Myr (top left panel), 63-251 Myr (top right) and $>$251 Myr (bottom left) along with super giant shells (black circle) in the LMC field.}
\label{clusters_sgs}
\end{figure}

In the last section, we demonstrated that the star clusters in the inner LMC are preferentially formed in the bar region.
The video-1 and video-2  suggest that the bar region of the LMC is clearly visible in the age range 60 - 250 Myr.
In Figure \ref{clusters_sgs}, we have plotted the distribution of clusters younger than 63 Myr, 63-251 Myr and $>$251 Myr, with all the groups put together. In the plots we have also shown the location of the Super giant shells \citep{kim99}. The younger clusters near the 30 Dor region are found to be located close to the shells.
As seen in the last sub-section, the bar region also has more of the poor and moderately rich clusters. The clusters older than 100 Myr, which is the peak of the cluster formation, is found to be located more towards the ends of the bar. On the other hand, around 100 Myr, we can detect the entire bar to be forming clusters. For younger ages, the clusters are formed preferentially in the central regions of the bar. Therefore, we clearly detect a propagation of cluster formation in the LMC bar, from the ends to the center of the bar. This is similar to that found by \citet{jacy2016} (OGLE IV cepheid) and \citet{piatti2015_vmcXVI}. On the other hand, using MACHO Cepheids data \citet{alcock99} found that there is a propagating star formation in the last 100 Myr, along the bar, from southeast to northwest. 

 Based on our results, we suggest that the bar of the LMC witnessed a burst in the cluster formation during the period 250 - 60 Myr, where the cluster formation started near the ends of the bar and then proceeded towards the centre of the bar. This could suggest that the bar of the LMC was active during this period and the direction of propagation of cluster formation could also support the idea that the bar
was effective in driving the gas towards the central region of the LMC during this period. Thus the LMC bar was active at least up to 60 Myr. 
We also detected a shrinking of cluster formation in the eastern side of the bar. Between the ages 100 Myr to 60 Myr, the cluster formation is clearly seen to be shrunk to inner regions in the eastern side. We speculate that this may be due to the compression of gas due to the movement of LMC in the Galactic halo. Detailed inspection is needed to verify and validate this claim. 

\onecolumn
\twocolumn

\subsection{Mass-Radius relation}
\begin{figure}
\includegraphics[width=\columnwidth]{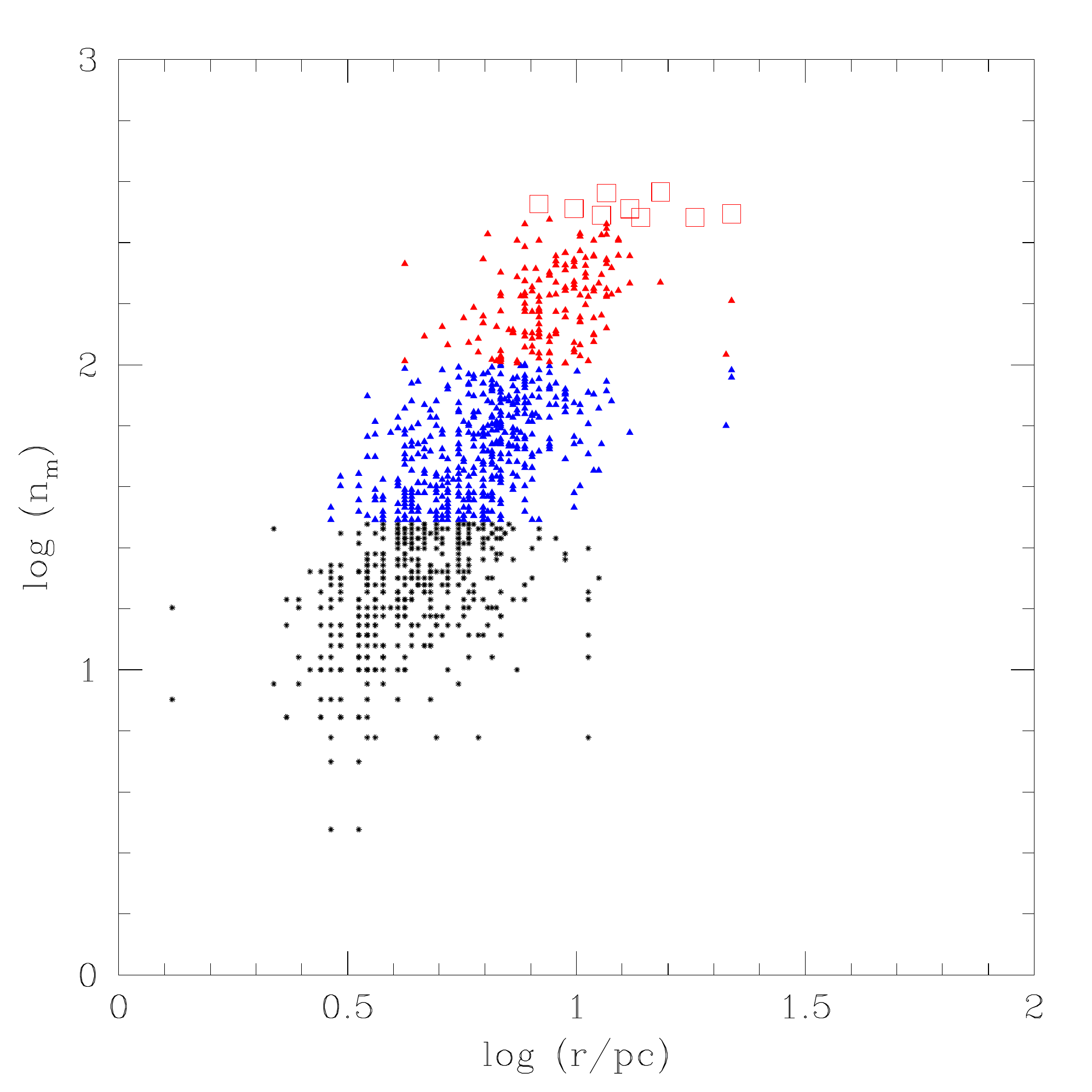}
\caption{Corelation between the radius and strength for different group of cluster. The point types used here are similar to Figure \ref{clusters_sgs}. }
\label{logr_lognm}
\end{figure}

\begin{figure}
\includegraphics[width=\columnwidth]{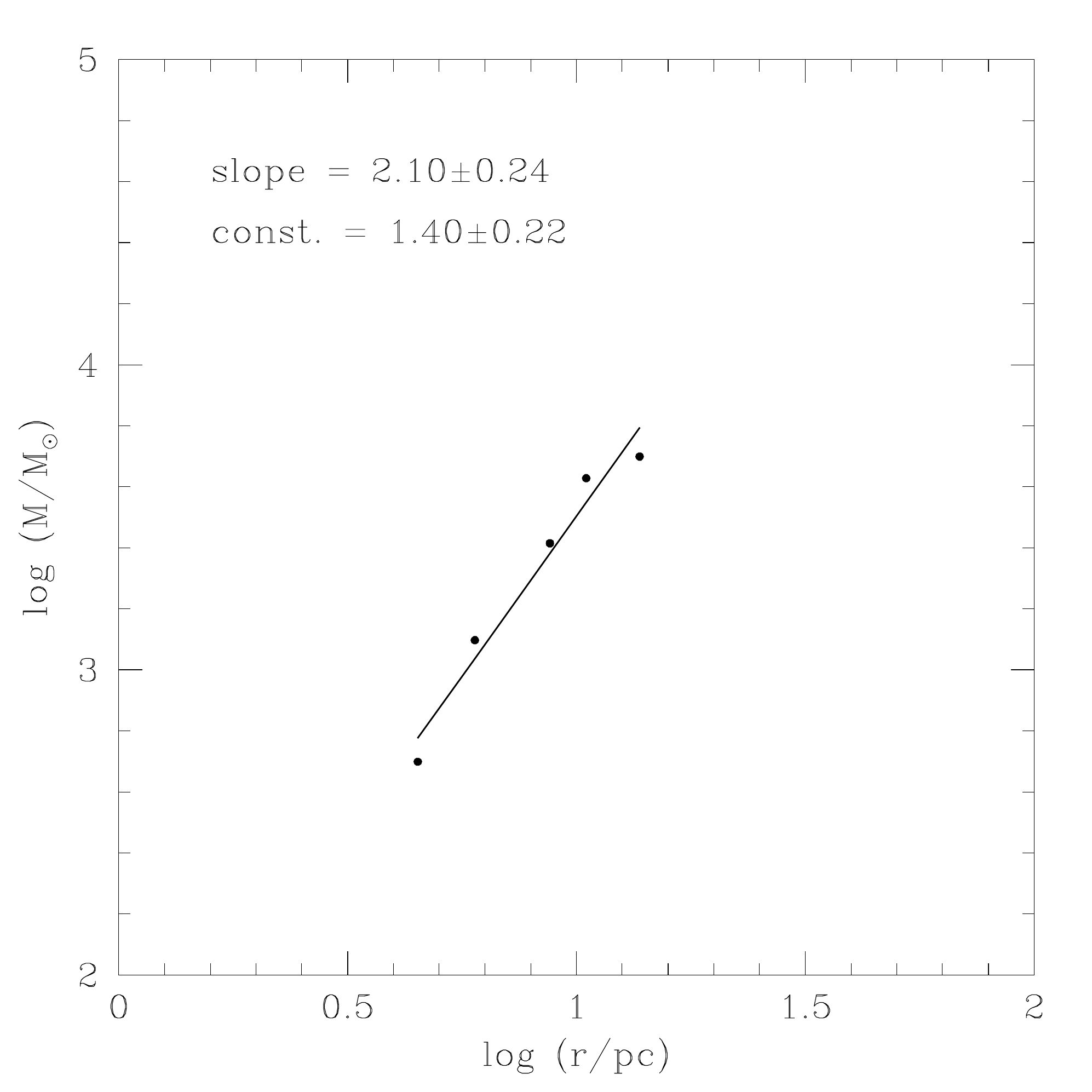}
\caption{Corelation between average radius of clusters and average mass of clusters of different groups . Slope and y-intercept of linear fit are also shown.}
\label{logr_logm}
\end{figure}

One of the fundamental relations concerning the structure of the clusters is the relation between the mass and radius of clusters. Since we have a large sample covering a range of strength, 
we tried to understand the correlation between the radius and strength clusters. We estimated the number of cluster members in each cluster,
which is taken as the proxy for the mass of the cluster. We plotted the radius (in pc) against the $n_m$ and found that it is a non-linear relation. On the other hand, we find that the correlation is logarithmic, as shown in Figure \ref{logr_lognm}. The four groups are shown in different colours and we see a continuous and similar variation across the groups.
This shows that there is a fundamental relation between the strength of the cluster and its radius. The strength of the cluster used here is the star count, on the other hand, it is best to use the mass of the cluster instead.  In order to obtain a first hand estimation of the slope, we took the average
value of radius for the 4 groups of clusters and used the average of their mass range. The linear fit to the data gives a slope of 
2.1 and a y-intercept of 1.4, as shown in Figure \ref{logr_logm}. A relation between the radius and mass
of star clusters in the solar neighbourhood was derived by \citet{pfal2016}. They used the relation M$_c$ =C${_m}{\times}R^{\gamma}$, and
estimated the value of gamma to be 1.71 for a large range of cluster mass. In our simplified method, we estimate the slope
to be 2.1. As mentioned by \citet{pfal2016}, it is necessary to find if there is a universal mass-radius
relation, and the effect of environment on this. In this study, we have attempted to estimate such a relation for the first
time, in the LMC, which has very different star forming properties, when compared to our Galaxy. We find that the value of gamma is not very different between the two galaxies. In a follow up study, we plan to estimate the mass of individual clusters and re-derive the above relation.

\subsection{Radius and age of clusters}
\begin{figure}
\includegraphics[width=\columnwidth]{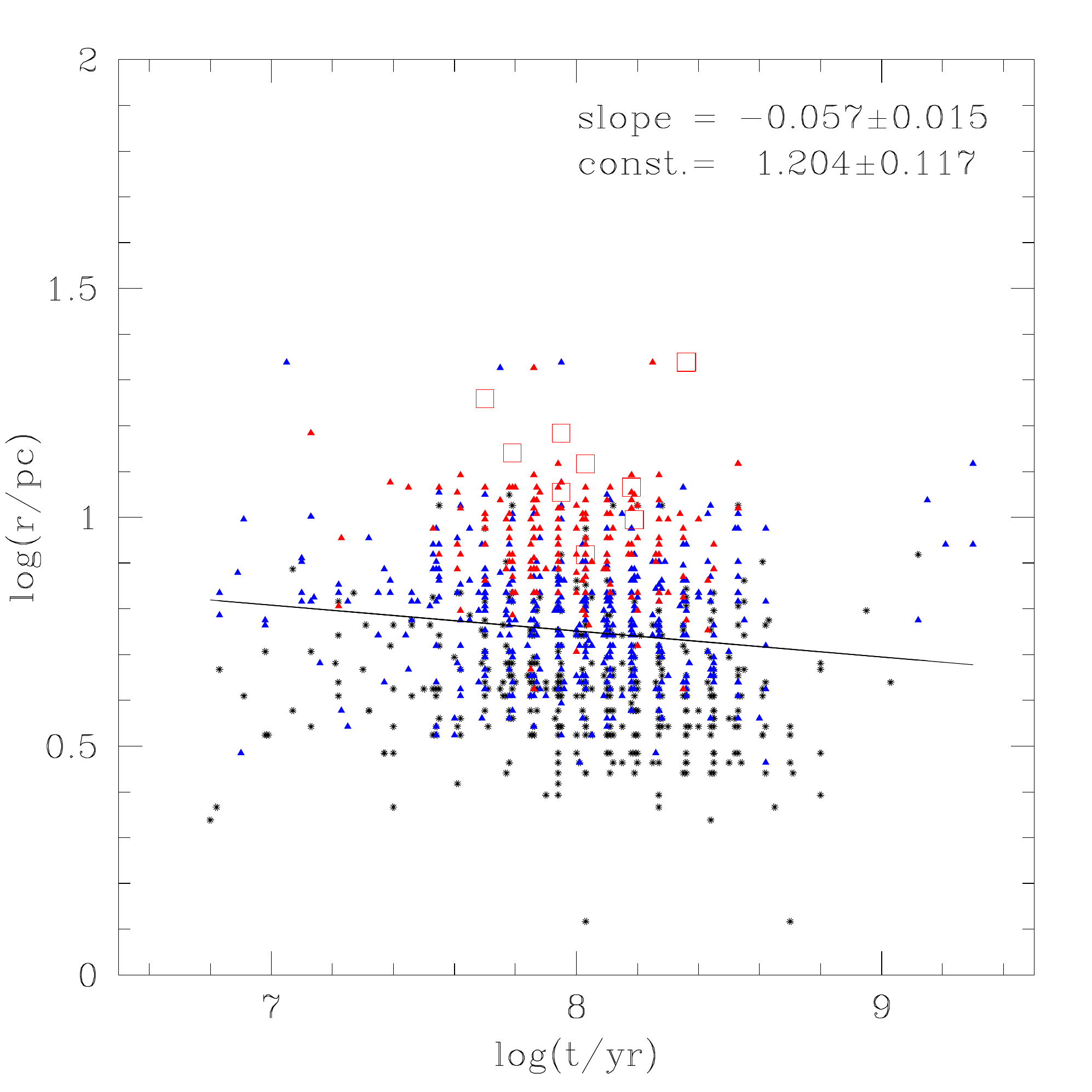}
\caption{The distribution of cluster radius with respect to its age. The distribution shows a linear relation with a slope of $-$0.057. The symbols used here are similar to Figure \ref{clusters_sgs}.}
\label{logr_logt}
\end{figure}

As the MCs have different environment and kinematics when compared to our Galaxy, it will be interesting to study the rate of dissolution
of star clusters. Such a study was presented by G10 and they did not find any significant difference in the size of the cluster
with age covered in their study. The sample of G10 had clusters of different mass and as the dissolution is a function of the
mass of the cluster, especially for the lower mass clusters, it is important to verify this for the various groups identified
in this study. As we have classified clusters as a function of strength, it will be interesting to find out whether there is
any change in the radius of the cluster as a function of age, for different groups. We have shown the log - log plot of age
against radius in Figure \ref{logr_logt}. The plots are fitted with straight line for different groups, and the coefficients ‘a’ and ‘b’ of the equation;
log(r) = a$\times$log(t)+b. The value of the coefficient, ‘a’, is found to be similar among all the groups, whereas the value of ‘b’, is found
to be different. This suggests that there is no significant difference between various groups, as far as change is radius as function of age is concerned. Therefore, we have shown the fit to the entire sample in the figure. As demonstrated we do not see any significant change in slope for the various groups, which
is similar to the result of G10. This might be due to the fact that the age range considered here is very narrow and hence we
might not be able to see any dissolution effect within this age range for any of the groups. Nevertheless, it is interesting
to notice that even the very poor group also did not show any indication of dissolution.

\subsection{Cluster age and reddening}
\begin{figure}
\includegraphics[width=\columnwidth]{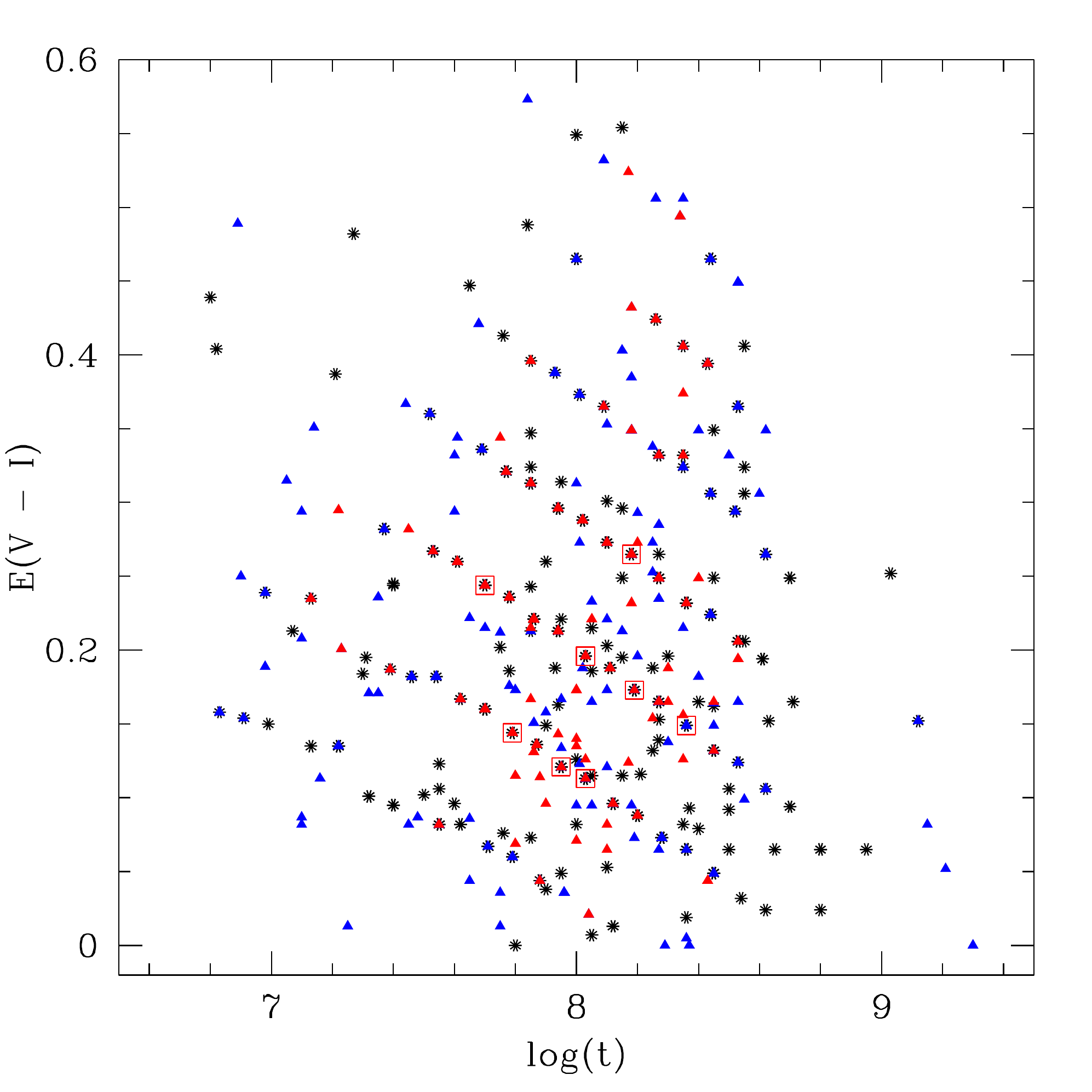}
\caption{The distribution of cluster reddening with respect to its age. The symbols used here are similar to Figure \ref{clusters_sgs}.}
\label{age_evi}
\end{figure}

It is well known that young star clusters have more reddening due to the presence of left over gas and dust from star formation around the cluster and  also, young clusters are found near star forming regions. Over a time, the left over material disappears and the reddening, in general, reduces with age of the cluster and also, the relatively older clusters are not found near star forming regions. It will be interesting to find the variation of reddening as a function of cluster age. Figure \ref{age_evi} shows the plot of age (log(t)) against reddening, E(V$-$I), for the four groups of clusters. As we can see, there is a range in reddening for a given value of age, which is due to the range of reddening estimated at various locations. We pay attention to the minimum reddening at a given age. The minimum reddening for a given age is found to decrease with age. The very poor and poor cluster have reddening $\sim$ 0.1 mag in E(V$-$I), whereas the moderate clusters have  slightly higher reddening, for an age of log(t) = 7.0. This reddening is found to decrease with age and found to reach the minimum value of reddening, by the age log(t) $\sim$ 7.6, for the very poor and poor  clusters. In the case of moderate clusters, the reddening reaches a minimum by log(t) $\sim$ 7.8. Thus, we find that the minimum reddening found for three groups of clusters is found to decrease with age and this might suggest that there is dispersal of left over material from the cluster with age. Our analysis suggests that it takes about 40 Myr for complete dispersal of material in very poor and poor clusters. In the case of moderate clusters, we find that it takes slightly longer, about 60 Myr to disperse the material. We are unable to comment on the rich clusters as the sample is small. A similar study has been done by \citet{sagar} with 15 young open clusters but did not find any uniformity in the relationship.

\section{Summary}

 We summarise the results of this study below:\\
1. We have classified and parameterised 1072 star clusters in the LMC using the OGLE III data and presented a catalog (full table available in the online version). The parameters of 308 clusters are presented for the first time.\\
2. We introduced a classification scheme for the star clusters in the LMC, based on their mass and demonstrated its usefulness.\\
3. We have introduced a semi-automated quantitative method to estimate ages of star clusters. \\
4. CMDs of 1072 clusters corrected for field star contamination and fitted with isochrones of estimated age and corrected for reddening 
   are made available in the online version.\\
5. Two videos which show the progression of cluster formation in the inner LMC as a function of age for various groups is available online.\\
6. Our study detects the peak of the cluster formation to be 125 $\pm$ 25 Myr for the very poor and poor clusters, whereas the moderate clusters are found to have the peak at 100 Myr.\\
7. The bar region of the LMC is found to be active in cluster formation during the period 60 - 250 Myr. We also suggest a progression of cluster
formation from the ends of the bar to the central region of the bar during the above period.\\
8. We find that the relation between the mass and radius of LMC clusters is similar to that found for the Galaxy.\\
9. We also demonstrate that the lower mass range of clusters is found in the LMC is very similar to the open clusters in the Galaxy.

\section{ACKNOWLEDGEMENTS}
We thank the referee for the valuable comments.

\onecolumn
\footnotesize
\begin{table}
\centering
\caption{A sample of full catalog is presented here. The table contains the name of the cluster, central coordinates (RA and Dec) as in B08, size of the cluster taken from B08,  estimated reddening and age in the columns 1-6 respectively. Column 7-10 represents the earlier estimations of ages by G10 (log(t$_{G10}$)), PU00 (log(t$_{PU00}$)), \citet{palma2016_catalog} (log(t$_{Pal16}$)) and \citet{piatti2014_vmcXII,piatti2015_vmcXVI} (log(t$_{P14,P15}$)). The last column represents the designated group number (I-V) of cluster.}
\label{Catalogue}
\begin{tabular}{lcccccccccc}
\hline
Star cluster & Ra & DEC & Radius & E(V$-$I) & log(t) & log(t$_{G10}$) & log(t$_{PU00}$) & log(t$_{Pal16}$) & log(t$_{P14,P15}$) & Group \\
 & (h m s) & ($\degr$ $\arcmin$ $\arcsec$) & ($\arcmin$) & & & & & & & \\
\hline
\hline
     KMHK15      & 4 36 20 &  -70 10 22  & 0.30  & 0.25  & 9.03  &  -    &     -      &     -     &  -   &    I  \\
     SL8         & 4 37 51 &  -69 01 45  & 0.75  & 0.08  & 9.15  & 8.80  &     -      &      9.20 &  -   &   II  \\
     SL12        & 4 39 59 &  -69 01 44  & 0.40  & 0.41  & 8.35  &  -    &     -      &     -     &  -   &    I  \\
     SL14        & 4 40 28 &  -69 38 57  & 0.55  & 0.12  & 7.95  & 8.30  &     -      &     -     &  -   &   II  \\
     SL15        & 4 40 43 &  -68 21 21  & 0.65  & 0.35  & 8.18  & 8.70  &     -      &     -     &  -   &   II  \\
     KMHK32      & 4 41 00 &  -68 18 04  & 0.29  & 0.19  & 8.61  &  -    &     -      &     -     &  -   &    I  \\
     BSDL2       & 4 41 09 &  -68 07 53  & 0.24  & 0.07  & 8.36  & 8.40  &     -      &     -     &  -   &    I  \\
     LW25        & 4 41 46 &  -68 10 18  & 0.43  & 0.25  & 8.27  &  -    &     -      &     -     &  -   &    I  \\
     LW28        & 4 42 10 &  -69 21 50  & 0.38  & 0.27  & 8.10  & 8.40  &     -      &     -     &  -   &    I  \\
     NGC1673     & 4 42 39 &  -69 49 12  & 0.49  & 0.24  & 7.70  & 7.70  &     -      &     -     &  -   &   II  \\
\hline
\end{tabular}
\end{table}

\normalsize
\twocolumn

\bibliographystyle{mn2e}

\label{lastpage}

\end{document}

%% file: journal_abbv.tex
\def\aj{AJ}%
\def\actaa{Acta Astron.}%
\def\apj{ApJ}%
\def\apjl{ApJ}%
\def\apjs{ApJS}%
\def\ao{Appl.~Opt.}%
\def\aap{A\&A}%
\def\aaps{A\&AS}%
\def\mnras{MNRAS}%
